\documentclass[%
 reprint, superscriptaddress,
 amsmath,amssymb,
 aps,
]{revtex4-2}

\usepackage{graphicx}
\usepackage{dcolumn}
\usepackage{upgreek}
\usepackage{xr-hyper}
\usepackage[hidelinks]{hyperref}
\usepackage{bm}
\usepackage{braket}
\usepackage{siunitx}
\usepackage{amsmath}

\begin{document}

% \preprint{APS/123-QED}

\title{Coherent manipulation of nuclear spins in the strong driving regime}

\author{Dan Yudilevich}
\affiliation{Department of Chemical and Biological Physics, Weizmann Institute of Science, Rehovot 7610001, Israel}

\author{Alon Salhov}
\affiliation{Racah Institute of Physics, The Hebrew University of Jerusalem, Jerusalem 9190401, Israel}

\author{Ido Schaefer}
\affiliation{Racah Institute of Physics, The Hebrew University of Jerusalem, Jerusalem 9190401, Israel}
\affiliation{Dahlem Center for Complex Quantum Systems and Fachbereich Physik, Freie Universit\"at Berlin, Arnimallee 14, 14195, Berlin, Germany}

\author{Konstantin Herb}
\affiliation{Department of Physics, ETH Z\"urich, Otto Stern Weg 1, 8093 Zurich, Switzerland}

\author{Alex Retzker}
\affiliation{Racah Institute of Physics, The Hebrew University of Jerusalem, Jerusalem 9190401, Israel}
\affiliation{AWS Center for Quantum Computing, Pasadena, CA 91125, USA}

\author{Amit Finkler}
\affiliation{Department of Chemical and Biological Physics, Weizmann Institute of Science, Rehovot 7610001, Israel}
\email{amit.finkler@weizmann.ac.il}

\date{\today}

\begin{abstract}
Spin-based quantum information processing makes extensive use of spin-state manipulation. This ranges from dynamical decoupling of nuclear spins in quantum sensing experiments to applying logical gates on qubits in a quantum processor. Fast manipulation of spin states is highly desirable for accelerating experiments, enhancing sensitivity, and applying elaborate pulse sequences. Strong driving using intense radio-frequency (RF) fields can, therefore, facilitate fast manipulation and enable broadband excitation of spin species.

In this work, we present an antenna for strong driving in quantum sensing experiments and theoretically address challenges of the strong driving regime. First, we designed and implemented a micron-scale planar spiral RF antenna capable of delivering intense fields to a sample. The planar antenna is tailored for quantum sensing experiments using the diamond's nitrogen-vacancy (NV) center and should be applicable to other solid-state defects. The antenna has a broad bandwidth of 22\,MHz, is compatible with scanning probes, and is suitable for cryogenic and ultrahigh vacuum conditions. We measure the magnetic field induced by the antenna and estimate a field-to-current ratio of $113 \pm 16$\,G/A, representing a six-fold increase in efficiency compared to the state-of-the-art, crucial for cryogenic experiments. We demonstrate the antenna by driving Rabi oscillations in \textsuperscript
{1}H spins of an organic sample on the diamond surface and measure \textsuperscript
{1}H  Rabi frequencies of over 500\,kHz, i.e., $\mathrm{\pi}$-pulses shorter than 1\,$\upmu\mathrm{s}$ -- faster than previously reported in NV-based nuclear magnetic resonance (NMR).

Finally, we discuss the implications of driving spins with a field tilted from the transverse plane in a regime where the driving amplitude is comparable to the spin-state splitting, such that the rotating wave approximation does not describe the dynamics well. We present a simple recipe to optimize pulse fidelity in this regime based on a phase and offset-shifted sine drive, which may be optimized \textit{in situ} without numerical optimization procedures or precise modeling of the experiment. We consider this approach in a range of driving amplitudes and show that it is particularly efficient in the case of a tilted driving field.

The results presented here constitute a foundation for implementing fast nuclear spin control in various systems.
\end{abstract}

\maketitle

\section{Introduction}
Quantum sensing with solid-state spin sensors, such as the nitrogen-vacancy (NV) center in diamond, frequently involves manipulating nuclear spin states. Nuclear spins may be part of the sample of interest, as in the case of nanoscale nuclear magnetic resonance (NMR) spectroscopy, which relies on sequences of radio-frequency (RF) pulses applied to the sample to recover information on its chemical structure~\cite{Mamin2013a, Loretz2014a, Aslam2017a, Abobeih2019b, Smits2019a}. Solid-state nuclear spins around the sensor are also utilized as ancilla qubits that store the quantum state of a sensor to retrieve it repeatedly~\cite{Lovchinsky2016, Aslam2017a} or to prolong the sensing time~\cite{Rosskopf2017}. 

Most experiments have relied so far on antennas that induce weak RF driving fields, with a standard $\pi$-pulse lasting a few tens of microseconds~\cite{Smeltzer2009, Mamin2013a, Lovchinsky2016}. These lengthy pulses imply longer measurement times and, thus, reduced sensitivity\,\cite{Degen2017}. They may also impede the application of elaborate pulse sequences, as the sensing time in NV-based NMR is limited by the spin relaxation time of the NV center ($T_1$)\cite{Laraoui2013a}, or a nuclear memory\,\cite{Aslam2017a}.

Fast manipulation of nuclear spins by strong RF driving fields can better utilize the limited sensing time of NV center sensors, generate broadband excitation of the nuclear spin resonance, and enable novel sensing protocols\,\cite{Munuera2023}. Previous works demonstrated strong driving for the NV center electron spin at a rate of $\sim 1\,\mathrm{GHz}$~\cite{Fuchs2009a}, and \textsuperscript{13}C spins in diamond at $\sim 70\,\mathrm{kHz}$~\cite{Herb2020}. The highest reported driving rates for protons in NV-based NMR are $ 50-80\,\mathrm{kHz}$~\cite{Mamin2013a, Aslam2017a}.

The system described below enables spin manipulation in a regime where the driving strength $\Omega_d$ is close to the energy splitting (i.e., $\Omega_d \lesssim \omega_0$). Alongside the experimental challenges of producing strong driving fields, working in this regime poses a theoretical control challenge. Most experimental setups deliver linearly polarized fields often tilted away from the transverse plane. For weak driving strengths ($\Omega_d \ll \omega_0$), the dynamics are accurately approximated by sinusoidal state transitions, known as Rabi oscillations, under the rotating-wave approximation (RWA). In the regime where $\Omega_d \lesssim \omega_0$, however, the deviations from an ideal drive (specifically, a linearly polarized drive with a longitudinal component) will markedly alter the dynamics~\cite{Fuchs2009a}. Without proper adaptations, this ``breakdown'' of the RWA results in the deterioration of pulse fidelity. It is thus crucial to design the signals so as to optimize an operation's fidelity in the strong driving regime. 

The issue of strong driving has attracted interest, especially for quantum information processing, where strong driving can accelerate operations and increase the speed of quantum processors~\cite{Song2016FastApproximation, Yang2017AchievingQubit}. Among others, optimal control strategies have been employed for optimizing quantum control in the strong driving regime. In particular, the concept of time-optimal control fields~\cite{Khaneja2001TimeSystems} has been introduced to identify the shortest possible signals to control a qubit state~\cite{Boscain2006TimeField, Wang2011Time-optimalDiamond}. Bang-bang control sequences have been shown to be the quickest form, while bang-bang driving at rates exceeding $\omega_0$ has been demonstrated on solid-state qubits~\cite{Avinadav2014Time-optimalDriving}, and other optimal control theory-derived waveforms have been demonstrated for solid-state qubit controls~\cite{Scheuer2014PreciseApproximation}. Optimal control approaches require a precise description of the driving field and the qubit, e.g., the relative orientation and magnitude, according to which the control signal is calculated. However, errors in the estimated parameters might deteriorate the ultimate performance compared with the simulations~\cite{Scheuer2014PreciseApproximation}. Also, \textit{in situ} optimization is difficult as it requires sampling a complex parameter space.

In this work, we design and implement a micrometer-scale planar spiral RF antenna compatible with NV magnetometry and capable of delivering intense RF pulses to a diamond sample. We characterize the antenna's characteristics and performance. Demonstrating the antenna's function by driving proton (\textsuperscript{1}H) spins, we observe spin state Rabi oscillations at frequencies surpassing 500\,kHz. 

We then discuss the unique characteristics of spin-state control by a strong and tilted driving field. We propose a novel approach to optimize the fidelity of control pulses in this regime, which is particularly suitable for driving fields that are noisy or not fully characterized.

\section{Experimental methods}
We performed experiments on a home-built room temperature confocal microscope. NV center electron spins were excited by a 520~nm diode laser, and their fluorescence was measured by a single-photon counting module. Low-frequency RF signals ($\sim$1\,MHz) were irradiated to the sample via our novel, custom-designed spiral antenna (see further in Sec.\,\ref{spiral_section}), and the NV center electronic spins were controlled by microwave pulses delivered by a wire drawn above the sample.

The diamond sample was a thin, single-crystal $\left[100\right]$ diamond membrane (approximately 30~$\upmu\mathrm{m}$ thick) patterned with nanopillar diamond waveguides. Shallow NV centers were created in the diamond by nitrogen ion implantation. For proton sensing, a small drop of microscope immersion oil (Cargille Type LDF) was applied to the diamond's surface with a sterile syringe (see further details in SI-2).

\section{Planar spiral radio-frequency antenna \label{spiral_section}}
\begin{figure}
\centering
\includegraphics[scale=0.25]{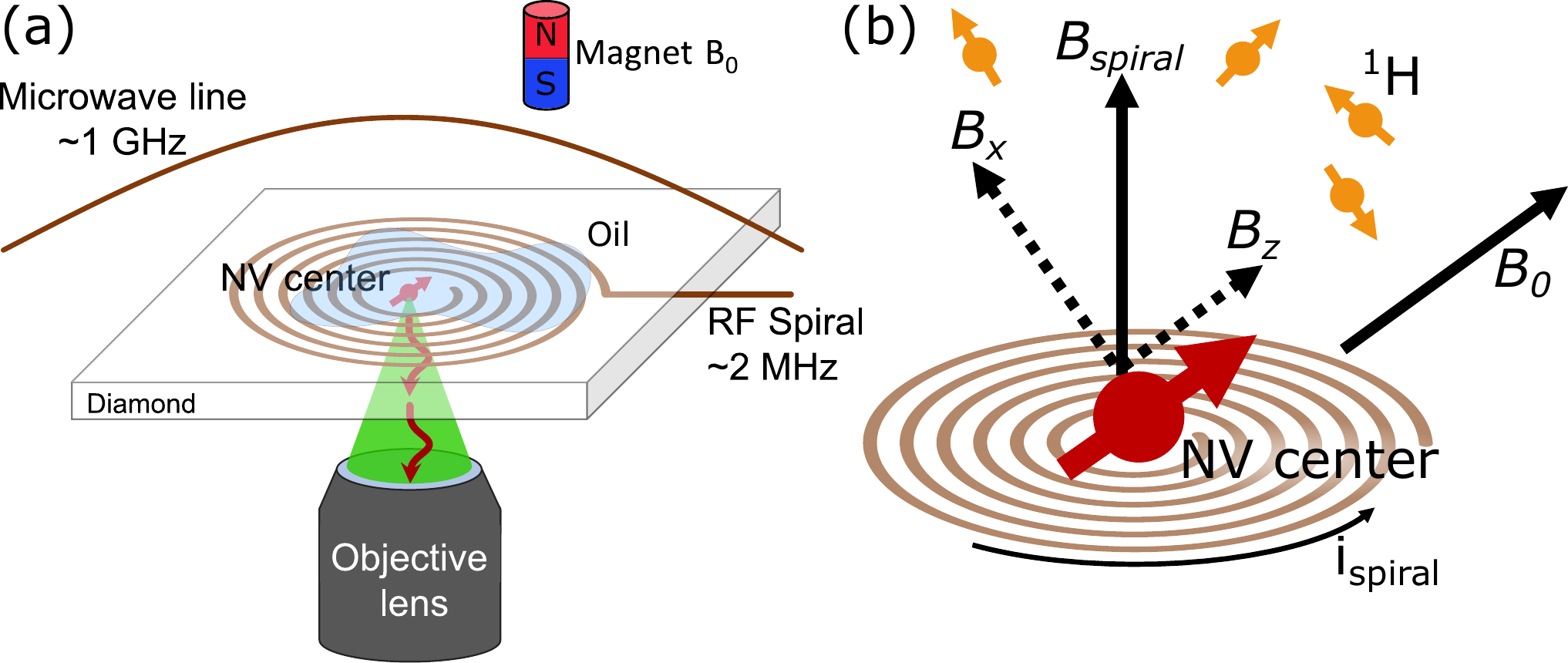}
\caption{\label{fig:schematic} Schematic of the experimental system (not to scale). (a) The RF spiral antenna sits underneath a thin diamond sample with NV centers in nanopillar waveguides. The NV centers are addressed optically through an aperture in the antenna. Microwave signals to the NV center are applied with a thin copper wire drawn above the diamond sample. Immersion oil with \textsuperscript{1}H is placed atop the diamond. (b) The static magnetic field $B_0$ is aligned with the NV center's axis. The spiral antenna's field is approximately perpendicular to the diamond's surface, inducing an RF field component along the NV's axis ($B_z$) and a component perpendicular to $B_0$ ($B_x$).}
\end{figure}

We designed the RF antenna as a planar spiral. Fig.~\ref{fig:schematic}(a) depicts the setup schematically. Compatibility with a typical NV magnetometry apparatus was the fundamental design principle. The diamond is placed directly on the antenna to enhance the magnetic field induced at the sample's position. A small aperture at the center of the spiral allows optical access to NV centers. An additional wire, drawn above the sample, is dedicated to signals in the gigahertz range for manipulating the NV center's electron spin.

The antenna was fabricated on a polyimide flexible printed circuit, suitable for ultra-high vacuum and cryogenic environments. The planar geometry can accommodate a scanning probe, such as an atomic force microscope, which may be used, for example, to carry a sample~\cite{Degen2009a} or create a magnetic field gradient~\cite{Grinolds2014}.

\begin{figure}
\centering
\includegraphics[width=1\columnwidth]{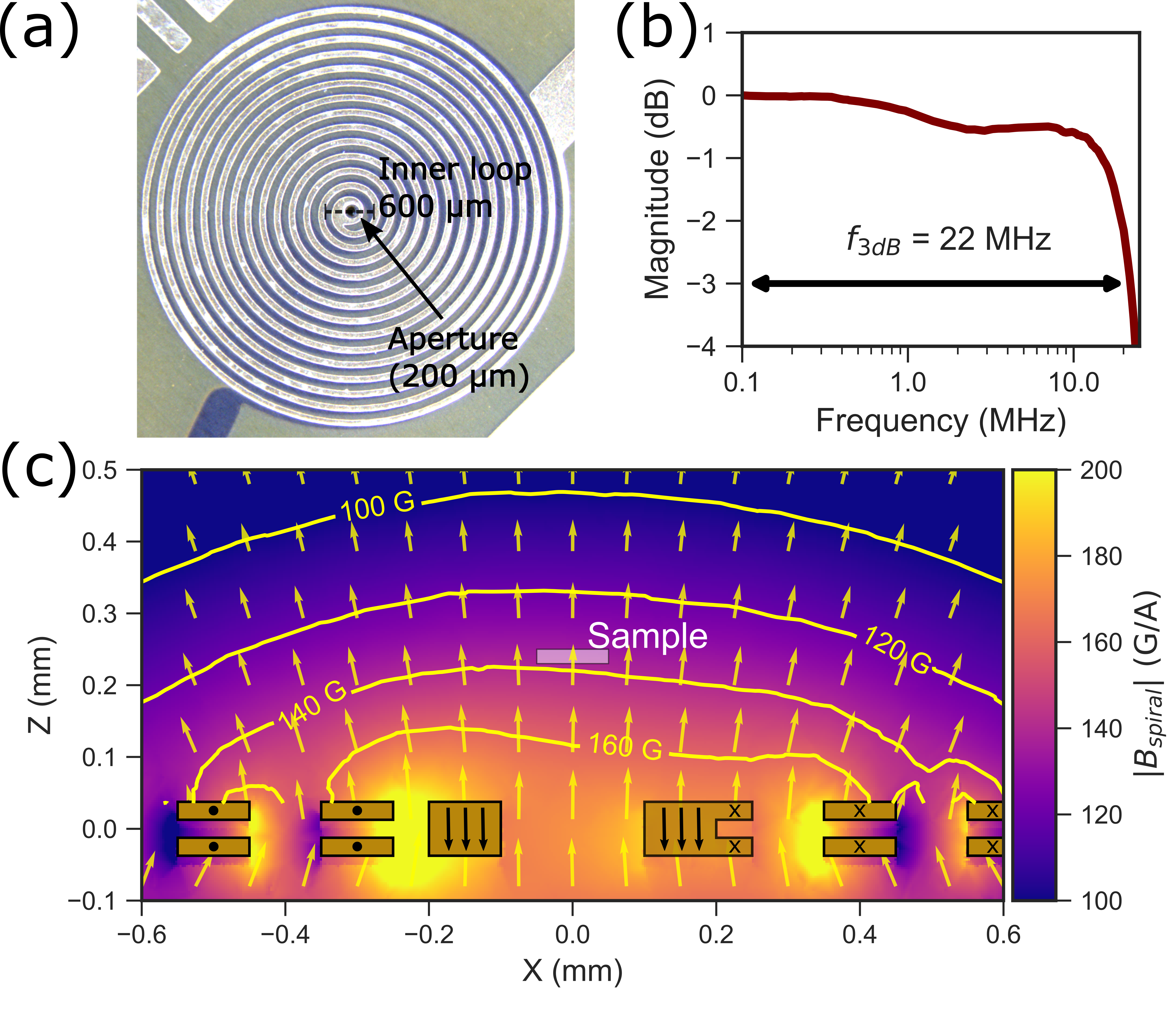}
\caption{\label{fig:antenna}Spiral broadband RF antenna. (a) Photo of the spiral antenna. The sample sits on the antenna, and the working region is directly above the aperture. (b) Transmission characteristics of the antenna ($S_{21}$ parameter). (c) Finite-element simulation of the antenna's field, focusing on the region of interest. The image shows a cross-section along the dashed line in (a). The color map and contours depict the magnetic field magnitude, and the arrows show the projection of the field's orientation on the XZ plane. The golden polygons depict the cross-section of the spiral's trace. The markings inside the polygons denote the direction of the simulated current. The sample's position in the current experiment is marked by the semitransparent rectangle.}
\end{figure}

In our antenna, the inner loop diameter is $600\,\upmu\mathrm{m}$, at the center of which is a $200\,\upmu\mathrm{m}$-diameter optical aperture; the sample's area of interest is placed on the aperture (Fig.~\ref{fig:antenna}(a)).  The inner diameter was minimized to achieve the strongest field up to the fabrication capabilities. The spiral consists of two identical layers, separated by a $20\,\upmu\mathrm{m}$ polyimide layer and connected by a via at the center. The number of turns and trace width of the spiral can be set to optimize its operation; for a larger field-per-current ratio, the number of turns should be increased. However, the bandwidth decreases with the number of turns, and the field-per-power is maximal for a specific number of turns (see further in SI-1). The results presented in this study were measured with a 15-turn spiral and a $100\,\upmu\mathrm{m}$ trace width. 

We terminate the antenna with a $\sim 50~\Omega$ load that dissipates over $90\%$ of the generated power. By monitoring the voltage on the load, we also determine the current through the antenna. The 3 dB bandwidth of the antenna is approximately 22 MHz, as observed in the transmission spectrum of the system ($S_{21}$ parameter, Fig.~\ref{fig:antenna}(b)). The antenna's bandwidth allows working with bias fields of up to 500 mT (for detecting proton spins). Such bias fields are required when wishing to utilize an ancilla nuclear spin in the diamond as a quantum memory~\cite{Neumann2010, Lovchinsky2016}. Additionally, the large bandwidth enables the transmission of pulses shorter than a microsecond without significant distortion.

Fig.\,\ref{fig:antenna}(c) shows a finite element simulation of the antenna's field distribution. The figure depicts the field along a cross-section of the antenna's center and where the sample sits. The simulation confirms that the expected magnetic field is approximately uniform in magnitude and orientation over the projected sample position. The magnetic field for a 1 A current at the experiment's sample position was estimated to be $136 \pm 1 ~\mathrm{G}$. The field vector was nearly perpendicular to the spiral plane, with a slight tilt of $1\pm0.7^{\circ}$.

\begin{figure}
\includegraphics[scale=0.65]{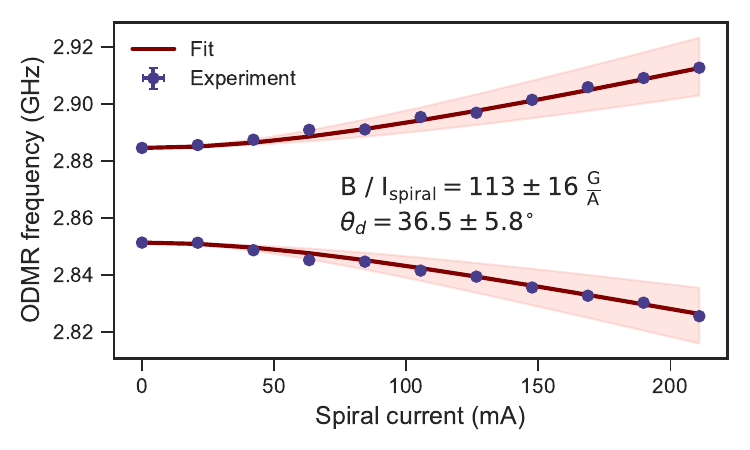}
\caption{\label{fig:dc_field_characterization}Direct current magnetic field characterization. The NV center level shifts were measured in a series of ODMR spectra with varying currents through the spiral. The points were fit to a model incorporating the magnetic field tilt and a strain field. The pink areas mark the confidence intervals of the fit.}
\end{figure}

We characterized the magnetic field vector emitted by the antenna using \textit{in situ} static magnetic field measurements with the NV center. We swept a direct current through the antenna and measured the Zeeman shift of the NV center's levels around $B_0 = 0$ (without an additional applied field). From the optically detected magnetic resonance (ODMR) spectra, we extracted the dependence of the field magnitude on the current and the tilt of the applied magnetic field to the NV axis. The result is plotted in Fig.~\ref{fig:dc_field_characterization}. We fit the data to a spin Hamiltonian incorporating strain and a magnetic field tilted away from the NV center axis. Thus, the transitions are not linearly dependent on the magnetic field (see SI-2 for further details on the analysis). From the transitions, we obtain the DC field-to-current ratio of $B~/~I_\mathrm{spiral}=113\pm16~\mathrm{\frac{G}{A}}$.  The field's angle is measured to be tilted from the plane transverse to the NV axis by $\theta_d = 36.5\pm5.8^{\circ}$(corresponding to a tilt of $\sim 1.2^{\circ}$ from the normal to the spiral plane). The measured field's magnitude agrees with the finite element simulation. The NV center lies at an angle of $\sim 54.7^{\circ}$ to the diamond surface, parallel to the spiral plane. Thus, the measured orientation is consistent with our expectation that the planar spiral antenna induces a field normal to its plane.

In a sensing experiment, as discussed in the following section, there will usually be an applied quantizing magnetic field ($\vec{B}_0$) along the NV center's axis ($\hat{z}$ in Fig.\,\ref{fig:schematic}(b)). Under the RWA, the transverse component ($B_x$ in Fig.\,\ref{fig:schematic}(b)) drives the spins and is proportional to the Rabi frequency. From the DC characterization, we estimate it to be $B_x~/~I_\mathrm{spiral}= B \cos \left( \theta_d \right) / I_\mathrm{spiral} = 92\pm14~\mathrm{\frac{G}{A}}$ (the magnitude might be attenuated according to the transmission at the specific frequency, as described in Fig.\,\ref{fig:antenna}(b)).

\section{Fast \textsuperscript{1}H Rabi oscillations}

\begin{figure*}
\centering
\includegraphics[width=1\textwidth]{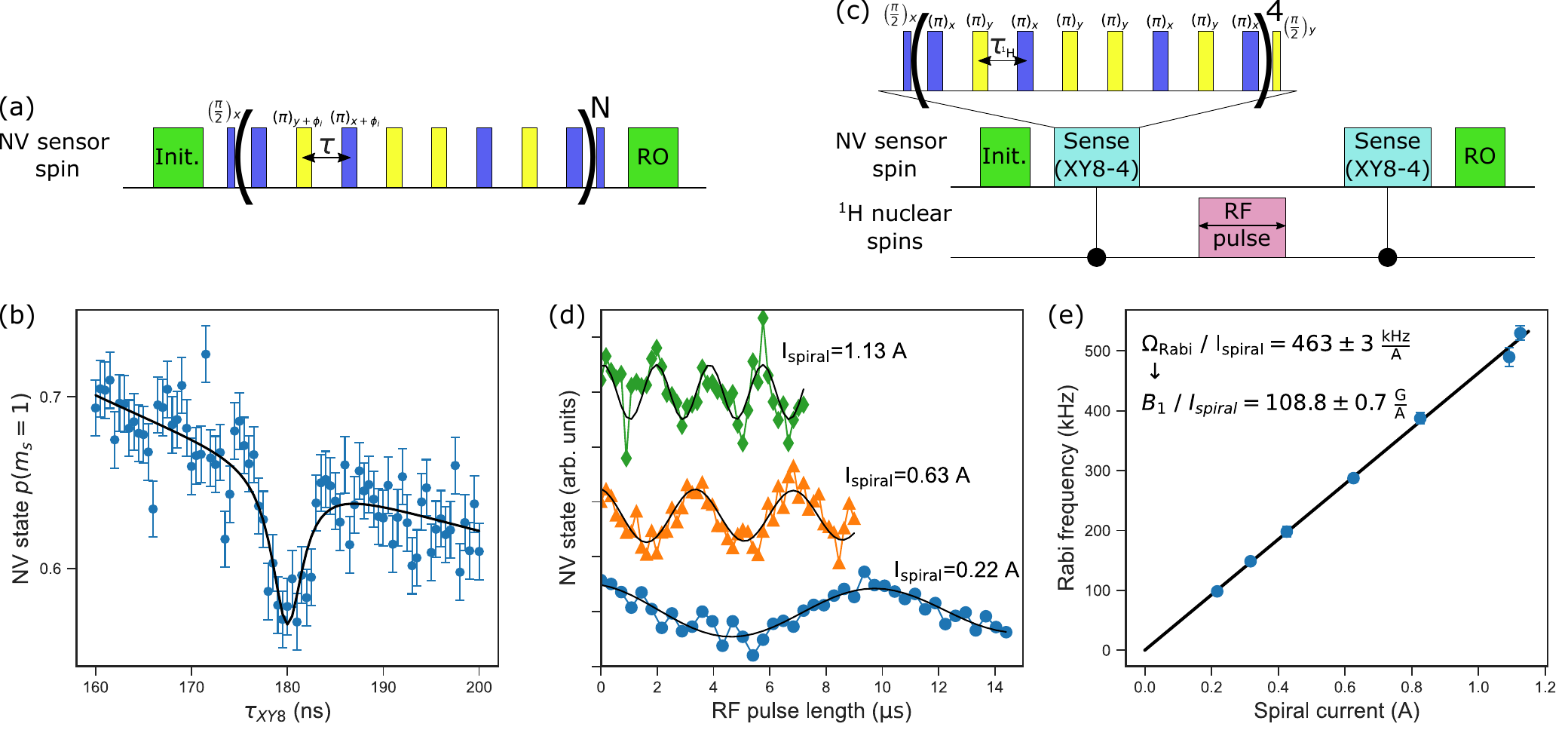}
\caption{\label{fig:proton_rabi} Fast Rabi oscillations of \textsuperscript{1}H nuclear spins. (a) A diagram of the randomized XY8-N pulse sequence used to sense the \textsuperscript{1}H nuclear magnetic resonance. (b) A randomized XY8-10 spectrum featuring a dip related to the \textsuperscript{1}H Larmor precession. (c) Diagram of the pulse sequence used to observe \textsuperscript{1}H nuclear spin Rabi oscillations. The nuclear spin precession was detected by correlating two XY8-4 dynamical decoupling blocks tuned to the \textsuperscript{1}H frequency found in (b). A varying radio frequency pulse tuned to the \textsuperscript{1}H frequency during the correlation time drives the \textsuperscript{1}H spin state. (d) Rabi oscillations of the \textsuperscript{1}H spins for different current amplitudes driven through the spiral antenna. (e) Summary of several Rabi frequencies measured, with a linear dependence on the current through the antenna.}
\end{figure*}

We demonstrate the antenna's function by driving Rabi oscillations in a proton spin ensemble of an organic sample on the diamond's surface. As a preliminary experiment, we detect proton nuclear magnetic resonance with an XY8-N dynamical decoupling sequence~\cite{Staudacher2013}, employing phase randomization to exclude spurious harmonics of \textsuperscript{13}C spins~\cite{Wang2019a}. Fig.\,\ref{fig:proton_rabi}(b) features an XY8-10 trace with a dip at the expected position of the proton Larmor frequency ($B_0\approx652~\mathrm{G},\ \omega_{^{1}\mathrm{H}}\approx 2.78~\mathrm{MHz}$), indicating that the NV senses the proton's oscillating magnetization.

We then employ a correlation spectroscopy sequence with RF pulses~\cite{Pfender2017a, Aslam2017a} to observe proton Rabi oscillations, as depicted in Fig.\,\ref{fig:proton_rabi}(c).  We use a correlation spectroscopy sequence based on XY8-4 dynamical decoupling blocks locked to the proton Larmor frequency to sense the phase of the proton's oscillation~\cite{Laraoui2015a, Staudacher2015a}. The correlation delay, i.e., the spacing between the two sensing blocks, was fixed at $20\,\mathrm{\upmu s}$. The RF pulses during the correlation delay, tuned to the proton Larmor frequency of varying duration, drive the nuclear magnetization, inducing a $\ket{\uparrow}\leftrightarrow\ket{\downarrow}$ transition.

The resulting Rabi oscillations are plotted in Fig.~\ref{fig:proton_rabi}(d) for several driving powers corresponding to different spiral currents. The oscillations were fitted to a decaying sine function, from which we extracted the driving frequency ($\Omega_d$). Fig.~\ref{fig:proton_rabi}(e) summarizes the observed driving frequencies as a function of spiral currents. The driving frequency is proportional to the driving current, as expected. We achieve a maximal driving frequency of $530\pm12~\mathrm{kHz}$, ultimately limited by the amplifier's saturation power.

We estimate a driving frequency-to-current ratio of $\Omega_d~/~I_\mathrm{spiral}=463\pm3~\mathrm{\frac{kHz}{A}}$. From this ratio, we estimate the field-to-current ratio of the transverse field at 2.78\,MHz to be $B_1~/~I_\mathrm{spiral}=108.8\pm0.7~\mathrm{\frac{G}{A}}$; this is in good agreement with the value expected from \textit{in situ} DC measurement ($92\pm14~\mathrm{\frac{G}{A}}$) presented previously and the finite-element simulations ($111.0 \pm 0.8 ~\mathrm{\frac{G}{A}}$).

\section{Manipulating spins in the strong driving regime} \label{sec:manipulating_strong_driving}

As our spiral antenna can indeed reach the strong driving regime ($\Omega_d \sim \omega_0/5$ in the aforementioned experiment), we describe a straightforward approach to generate control signals in the $\Omega_d \lesssim \omega_0$ regime for high-fidelity operations. We show that a simple sine signal with an offset may provide sufficient fidelity in this regime by optimizing just one or two parameters. 

Our approach is particularly suitable for tilted drive signals, that is, signals with a component along the quantization axis (hereinafter referred to as $\hat{z}$). Tilted drives are found in various solid-state spin qubit systems, such as the NV center in diamond as described in the previous section, the SiV defect in diamond~\cite{Nguyen2019}, defects in SiC~\cite{miao2020universal} and in h-BN~\cite{Stern2022Room-temperatureNitride}, as well as superconducting flux qubits~\cite{Yan2013Rotating-frameEvolution}. However, to our knowledge, optimizing strong tilted drives has not been discussed in the literature.

In what follows, we motivate our approach analytically using a clear physical picture, illustrate its validity numerically, and compare it to optimal control-derived signals. We argue that offset-sine signals bear benefits over optimal control-derived signals while providing similar and sufficiently high fidelity rates. Our focus is on the optimization of the $\pi$-pulse, which is to be reached at $t_{\pi} \sim \frac{\pi}{\Omega_d}$ (a precise definition follows below).

\subsection{Resonant offset-sine driving pulses}

We consider a two-level system driven by a tilted driving field. The system is described by the following Hamiltonian:
\begin{equation} \label{eq:Hd}
\mathcal{H} = \frac{\omega_0}{2} \sigma_z + \Omega_d f \left( t \right) \left(\sigma_x + \tan \left( \theta_d \right) \sigma_z \right) 
\end{equation}
where $\omega_0$ is the energy splitting of the two-level system, $\Omega_d$ is the maximum driving field amplitude, $\left| f(t) \right| \leq 1$ is the waveform, and $\theta_d \in \left[ 0, \frac{\pi}{2} \right)$ is the driving field's tilt angle from $\hat{x}$. Under this definition, the drive vector is not normalized; rather, the field's magnitude depends on the angle $\theta_d$.

In the weak-driving regime, conventional driving pulses are based on resonant sine waveforms, i.e., $f(t) = \sin \left( \omega_0 t + \varphi_d \right)$. The standard analysis proceeds with the rotating-wave-approximation (RWA)~\cite{Slichter1990}, which neglects the $\hat{z}$ component of the drive (i.e., assuming $\theta_d=0$, see Fig.\,S2(a) for schematic), as well as the counter-rotating term of the transversal component. The resulting rotating frame Hamiltonian is $\mathcal{H}_I = \frac{\Omega_d}{2}(\sin(\varphi_d)\sigma_x-\cos(\varphi_d)\sigma_y)$. In this regime, the only effect of the phase, $\varphi_d$, is to determine the axis of Rabi nutation, and for a $\pi$ pulse, the effect vanishes. 

In the strong-driving regime, we expect the dynamics to depend on $\varphi_d$ beyond the trivial dependence of the weak-driving regime. This is based on the observation that finite-duration sine waveforms contain a DC component. Namely, the zero-frequency Fourier component, $\frac{2\Omega_d}{\pi} \int_{0}^{\pi / \Omega_d} \sin \left(\omega_0 t + \varphi_d \right) \mathrm{dt}$, which depends on $\varphi_d$, is significant for $\Omega_d / \omega_0 \sim 1$. Thus, in this regime, we anticipate that varying the phase modulates the interplay between the different terms of the Hamiltonian, offering flexibility for pulse fidelity optimization.

We suggest utilizing the phase $\varphi_d$ to mitigate the effects of the counter-rotating term in the regime of $\Omega_d \lesssim \omega_0$ and optimize pulse fidelity rates in this regime. This may be supplemented by a DC offset to the drive, serving as another DC component that may be controlled to optimize the pulse. We note that the phase of the driving field was shown to be important in the strong driving regime in NMR already more than five decades ago~\cite{MacLaughlin1970SuppressionResonance}. More recently, the phase's effect was shown in single solid-state qubit experiments~\cite{London2014StrongFields, Rao2017NonlinearApproximation} and in NMR~\cite{Bidinosti2022GeneratingApproximation}. However, the phase has not been discussed in the context of tilted drives or in combination with a DC offset.

The first-order correction to the rotating frame Hamiltonian $\mathcal{H}_I$ is the Bloch-Siegert shift \cite{Bloch1940MagneticFields} that acts as an effective DC field along $\hat{z}$ in the rotating frame~\cite{Zeuch2020ExactApproximation}. Thus, the DC component of the longitudinal driving field (present for tilted field $\theta_d > 0$) may assist in canceling out the effects of the counter-rotating term. Let us now consider the special case of driving at an amplitude of $\Omega_d = \frac{\omega_0}{2 \tan \left( \theta_d \right)}$. In this case, a constant (DC) waveform equal to $-1$ yields an ideal driving Hamiltonian $\mathcal{H} = \Omega_d \sigma_x$. These observations motivate us to consider waveforms $f(t)$ based on the ``offset-sine'' waveform:
\begin{equation} \label{eq:Omega_d}
    f(t) \equiv \epsilon \left( t \right) \left(a +\left(1 - \left| a \right| \right)  \sin \left( \omega_0 t + \varphi_d \right) \right) 
\end{equation}
where the optimization parameters are $\left| a \right| \leq 1$ (the DC offset component) and $\varphi_d$ (the phase). For $a=0$, we obtain a standard sine (symmetric around 0), while for $\left| a \right|=1$ we get a constant DC drive.

In Eq.\,\ref{eq:Omega_d} we introduced the pulse's envelope function $0 \leq \epsilon \left( t \right) \leq 1$, which is zero at the pulse edges ($\epsilon \left( t_0 \right) = \epsilon \left( t_\mathrm{pulse} \right) = 0$). For weak driving, a simple rectangle function is often used as the envelope (i.e., rectangular pulse shape). However, as realistic transmission lines always have limited bandwidth, a discontinuous $\epsilon \left(t \right)$ will result in a distorted signal, and this distortion is significant for strong and short pulses. A smooth envelope function with finite rise and fall times can fit the signal into a prescribed bandwidth~\cite{Vandersypen2005NMRComputation}, and here specifically, we used an error-function pulse envelope~\cite{Bradley2019AMinute} (see Eq.\,S.2 and Fig.\,S2(b) in the SI for a schematic pulse).

\subsection{Optimizing control pulses based on the offset-sine waveform}

We demonstrate the performance of the offset-sine waveforms by numerically calculating the state evolution of a qubit under such a drive Hamiltonian (Eq.\,\ref{eq:Hd} and Eq.\,\ref{eq:Omega_d}). We consider as examples driving amplitudes of  $\Omega_d = \frac{\omega_0}{10}, \frac{\omega_0}{3}, \omega_0$. We focus on $\pi$-pulses, i.e., flipping the initial state $\ket{\uparrow}$ with the goal of maximizing the probability for $\ket{\downarrow}$. 

As an illustration, we choose parameters inspired by the aforementioned spiral antenna, namely, a drive tilt of $\theta_d = 35.3^{\circ}$, and limit the signals to a bandwidth of $\lesssim 10 \omega_0$ using an error-function envelope with a rise-time $\delta t = \frac{\pi}{10 \omega_0}$. The pulse durations are extrapolated from the weak driving regime and set to be $t_{\pi} = \frac{\pi}{\Omega_d} + 2 \delta t$, which accounts for the rise and fall times of the signal (for further details, see SI-3). 

\begin{figure*}
\centering
\includegraphics[width=1\textwidth]{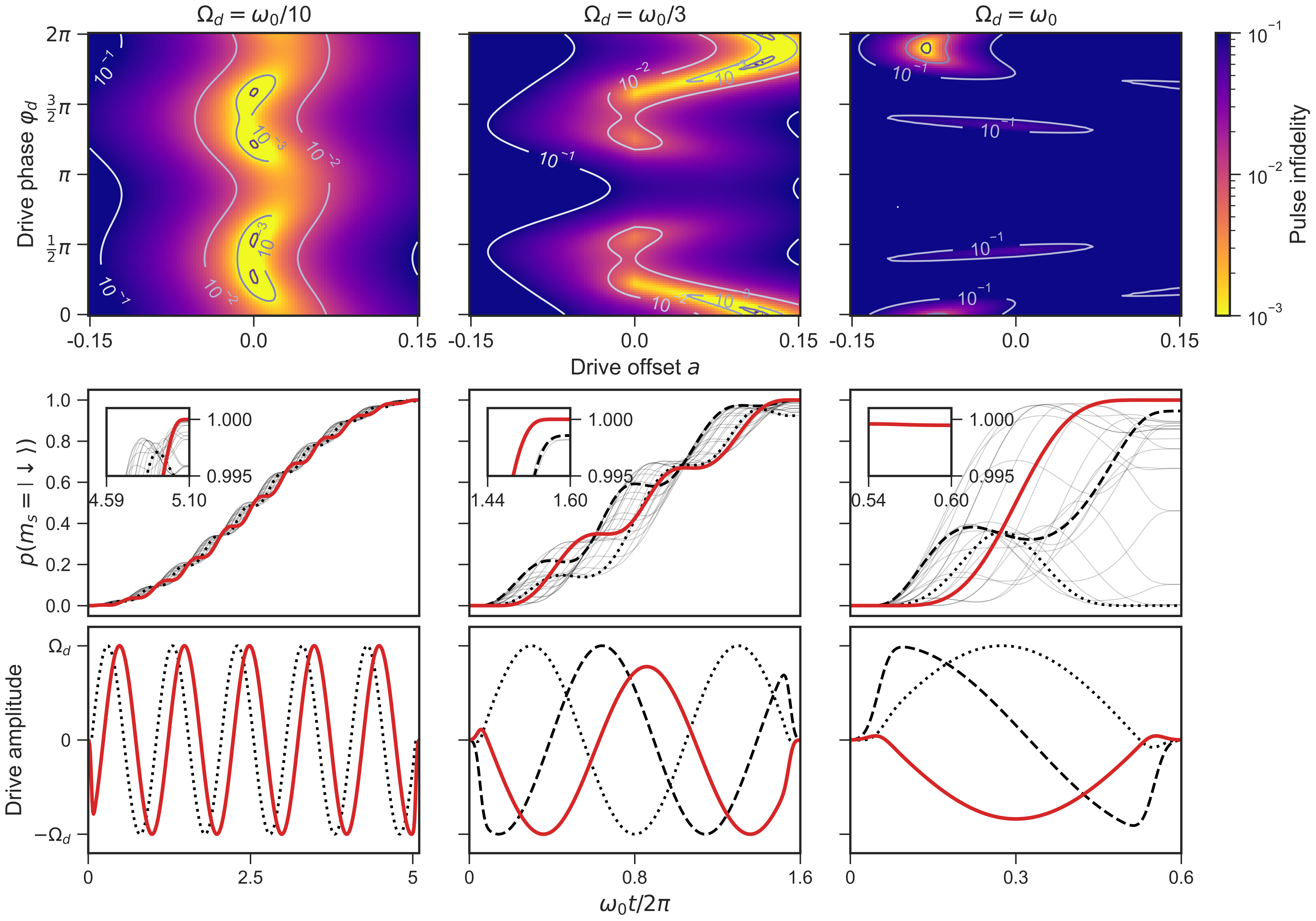}
\caption{\label{fig:offset-sine-driving} Optimizing offset sine $\pi$-pulse drives at different driving strengths. Top row: Calculating pulse infidelity for several driving strengths $\Omega_d$ as a function of pulse drive phase $\varphi_d$ and offset $a$.
Center row: spin state evolution for pulses at the corresponding drive strengths. Trajectories for many values of $\varphi_d$ at $a=0$ are shown in light curves. The best (worst) $\varphi_d$ values for $a=0$ are marked by dashed (dotted) curves. Evolution for pulses with both optimal phase and offset are shown by solid red curves. Insets focus on the pulses’ ends to highlight the final fidelity for each case.
Bottom row: the waveforms corresponding to the trajectories drawn in the center row, with matching curve format.}
\end{figure*}

We numerically calculate the pulse fidelity according to $\mathcal{F}=\left| \braket{\psi \left( t_{\pi} \right) | \downarrow} \right| ^2$ under the driving field for each driving amplitude, sampling various values of the phase ($\varphi_d$) and offset ($a$) of the signal. The results are presented in Fig.\,\ref{fig:offset-sine-driving} (top row) in terms of infidelity $1 - \mathcal{F}$ to contrast the results. Fig.\,\ref{fig:offset-sine-driving} (center row) shows the state evolution for various driving signals at the driving amplitude of the corresponding column. Evolutions are shown for various phases at zero-offset ($a=0$, light gray curves), with the zero-offset phase yielding the best (worst) pulse fidelity marked by dashed (dotted) curves. Evolutions under an optimal offset-sine drive are marked by red curves. The optimal offset-sines have offset and phase corresponding to the coordinates of minimum infidelity in the diagrams of the top row, i.e., the brightest points.  The bottom row shows the waveforms corresponding to the different state evolutions in the center row.

The center row of Fig.~\ref{fig:offset-sine-driving} illustrates how increasing the driving strength $\Omega_d$ from $\frac{\omega_0}{10}$ to $\omega_0$ leads to increasing deviation from the standard sinusoidal evolution characteristic of the RWA. For the stronger drive amplitudes, adjusting the drive phase $\varphi_d$ is crucial: for the extreme case of $\Omega_d = \omega_0$, a correct choice of drive phase $\varphi_d$ will yield $\mathcal{F} \approx 0.94$, while the worst choice will yield $\mathcal{F} \approx 0$.  Additional optimization of the DC offset significantly impacts the final state fidelity for the strongest drive amplitudes. For example, at $\Omega_d\lesssim \omega_0$, adding a proper offset will increase the fidelity to $\mathcal{F} > 0.999$, beyond the fault-tolerance threshold for some quantum computer architectures~\cite{Fowler2012SurfaceComputation, Barends2014SuperconductingTolerance}.

The drive phase is a single optimization parameter to optimize strong drive pulses and obtain pulse fidelities over 0.9, sufficient for quantum sensing tasks. The optimal choice of phase would depend on the driving amplitude and envelope function~\cite{Zeuch2020ExactApproximation}. 

The DC offset we introduced as a novel optimization parameter may also be significant, particularly for a driving field tilted by $\theta_d$ from $\hat{z}$. The drive tilt even serves as an additional resource: a tilted drive can achieve higher fidelities than driving fields purely along $\hat{x}$ when both the phase and offset are optimized (see Fig.\,S3(a) in the SI).

\subsection{Comparing with optimal control theory signals}

We compare our strategy with control signals generated by quantum optimal control theory (OCT) \cite{Rabitz,k67,k193,gross07,Ido_thesis}. In OCT, the optimization task is formulated as a maximization problem of a functional object by means of variational calculus. This yields a set of control equations, which are solved numerically by optimization algorithms. The optimization target is the maximization of the occupation of $\ket{\downarrow}$ at a predefined final time, $t_{\pi}=\frac{\pi}{\Omega_d}+2\delta t$, as defined previously. 

For our experiment, additional restrictions were added to the control problem:
\begin{enumerate}
    \item A restriction was imposed on the total energy of the drive that effectively kept the peak amplitude near $\Omega_d$. However, we emphasize that in the OCT solution, there was no explicit restriction on the amplitude as in the offset-sine optimization. As a result, some OCT waveforms shown below exceed an amplitude of $\Omega_d$.
    \item The spectral composition of the drive was restricted to $<10.7\omega_0$ to produce an experimentally realistic waveform with a smooth temporal profile. As mentioned before, this value is inspired by the spiral antenna.
    \item Homogeneous boundary conditions were imposed on the drive and its temporal derivative (i.e., zero drive and zero time-derivative of the drive at $t=0$ and $t=t_{\pi}$). This was done in order to obtain a realistic, smooth rise and fall of the drive.
\end{enumerate}

The optimization problem was solved for six values of $\Omega_d$, corresponding to different values of $t_{\pi}$. The drive tilt was set to $\theta_d = 35.3^{\circ}$. We compared the OCT signals with two forms of offset-sine signals: an approximation to the OCT signal, obtained by a least-square fitting of the OCT signal to Eq.\,\ref{eq:Omega_d}; and the optimal offset-sine, obtained by optimizing the offset and phase for the same parameters as the OCT signals.  The offset and phase optimization was done, as previously described, by fixing $t_{\pi}$  and the amplitude $\Omega_d$, and sampling a range of offsets and phases, choosing the offset and phase set that minimizes the infidelity. 

\begin{figure*}
\centering
\includegraphics[width=1\textwidth]{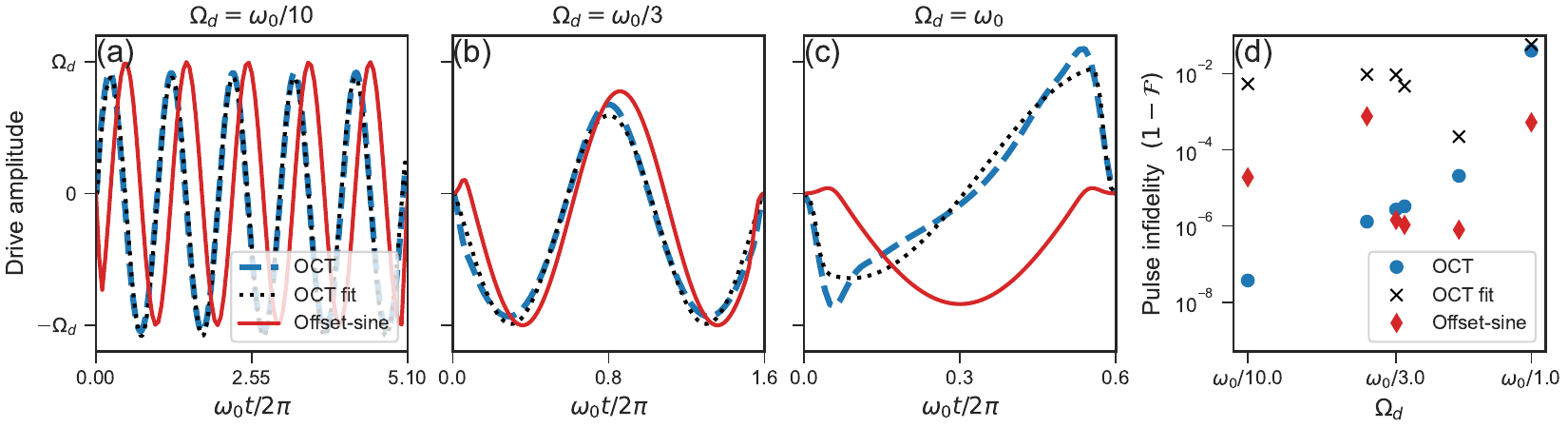}
\caption{\label{fig:optimal_control} Comparing OCT drive signals with offset sine drives. (a)-(c) Drive signals for different drive strengths $\Omega_d$ denoted above the plots, corresponding to different pulse durations $t_{\pi}=\frac{\pi}{\Omega_d} + 2\delta t$. The blue dashed curves are OCT waveforms. The dotted black curves are approximations of the OCT waveform by an offset-sine obtained by least-square fitting (OCT fit). The solid red curves are optimized offset-sine waveforms. (d) Infidelity rates of the driving signals for each of the shown waveforms.}
\end{figure*}

The signals for three cases are presented in Fig.\,\ref{fig:optimal_control}(a)-(c). The fitted offset-sine approximates the OCT-generated signals well, supporting the offset-sine approach. Conversely, the optimal offset-sine is very similar when  $\Omega_d \ll \omega_0$, but takes on a distinct shape as $\Omega_d$ approaches $\omega_0$. The difference, however, does not come at the expense of fidelity.

Fig.\,\ref{fig:optimal_control}(d) compares the pulse fidelity rates for the OCT signals, the approximated offset-sines, and the optimized offset-sine (for clarity, the data is presented in terms of infidelity $1 - \mathcal{F}$). The optimized offset-sine signals differ from the OCT signals but provide comparable fidelity rates, with $\mathcal{F}>0.999$. Interestingly, for the highest values of $\Omega_d$ considered here (i.e., shortest $t_{\pi}$ times), the fidelity rate of the optimized offset-sine even surpasses that of the OCT signal. While the difference presumably stems from the details of the optimization procedure, it underlines the potential of the optimized offset-sine waveforms as an alternative optimization strategy.

Our strategy for controlling qubits in the regime of  $\Omega_d \lesssim \omega_0$ thus relies on optimizing an offset-sine driving signal as an alternative to existing approaches for designing strong driving pulses, namely optimal control theory~\cite{Scheuer2014PreciseApproximation} and bang-bang control sequences~\cite{Boscain2006TimeField}.  The optimized offset-sine approach does not require precise driving field characterization and prior numerical optimization, and it suits a tilted driving field. Although the fidelity rates of the OCT and optimized offset-sine waveforms  differ, the real rate would likely be lower due to deviations between the simulated and actual drive parameters. This emphasizes the benefit of a strategy that conveniently enables \textit{in situ} optimization. The offset-sine signal may be optimized experimentally by varying over one or two parameters, namely the drive phase and DC offset. As such, this strategy is convenient to minimize deviations between the real and simulated conditions, for example, due to driving noise or a limited bandwidth~\cite{Scheuer2014PreciseApproximation}. 

\section{Summary and conclusions}

We developed a broadband spiral antenna tailored for quantum sensing experiments with NV centers, such as nanoscale NMR. The antenna's bandwidth suits nuclear spins at fields up to $0.5~\mathrm{T}$. We drive \textsuperscript{1}H spins at a Rabi frequency of over 500 kHz, faster than previously reported. The field-to-current ratio of the spiral antenna is three-fold better than the state-of-the-art, and the field-to-power ratio is over ten-fold better \cite{Herb2020}. Thus, owing to a low field-to-current ratio, it is possible to drive spins at appreciable driving frequencies with low power consumption, e.g., a Rabi frequency of over 100~kHz requires less than 2.5~W input power, making it especially appropriate for sensitive samples or cryogenic environments.

Furthermore, we discussed the issue of driving spins in a strong driving regime where $\Omega_d\lesssim \omega_0$. We show that spins may be flipped with high fidelity by utilizing resonant offset-sine drive pulses optimized by varying the drive field's phase and offset. Our approach obtains fidelity rates comparable to optimal control-derived signals and can be conveniently optimized \textit{in situ},  which is significant in experimental settings where the driving field is noisy or not fully characterized. Also, offset-sine signals are especially suitable for tilted driving fields. Pulse fidelities over 0.95 may be achieved by optimizing the drive phase, while varying the offset may bring fidelity rates over 0.999, above the fault-tolerance threshold.

\begin{acknowledgments}
We thank Nicolas Staudenmaier and Nabeel Aslam for delightful discussions on correlation spectroscopy. We thank Yonatan Vernik and Leah Fuhrman Javitt for their contributions. A.S.\, gratefully acknowledges the support of the Clore Israel Foundation Scholars Programme, the Israeli Council for Higher Education, and the Milner Foundation. I.S.\,acknowledges financial support by the German Federal Ministry of Education and Research (BMBF), project no.\ 13N15929 QCStack. A.F.\,is the incumbent of the Elaine Blond Career Development Chair in Perpetuity and acknowledges support from the Israel Science Foundation (ISF grants 963/19 and 419/20) as well as the Abramson Family Center for Young Scientists, the Willner Family Leadership Institute for the Weizmann Institute of Science and the Helen and Martin Kimmel Institute for Magnetic Resonance Research. We are grateful for the historic generosity of the Harold Perlman Family. A.R. acknowledges the support of ERC grant QRES, project number 770929, Quantera grant MfQDS, ISF and the Schwartzmann university chair. 
\end{acknowledgments}

\bibliography{references_spiral}

\begin{thebibliography}{48}%
\makeatletter
\providecommand \@ifxundefined [1]{%
 \@ifx{#1\undefined}
}%
\providecommand \@ifnum [1]{%
 \ifnum #1\expandafter \@firstoftwo
 \else \expandafter \@secondoftwo
 \fi
}%
\providecommand \@ifx [1]{%
 \ifx #1\expandafter \@firstoftwo
 \else \expandafter \@secondoftwo
 \fi
}%
\providecommand \natexlab [1]{#1}%
\providecommand \enquote  [1]{``#1''}%
\providecommand \bibnamefont  [1]{#1}%
\providecommand \bibfnamefont [1]{#1}%
\providecommand \citenamefont [1]{#1}%
\providecommand \href@noop [0]{\@secondoftwo}%
\providecommand \href [0]{\begingroup \@sanitize@url \@href}%
\providecommand \@href[1]{\@@startlink{#1}\@@href}%
\providecommand \@@href[1]{\endgroup#1\@@endlink}%
\providecommand \@sanitize@url [0]{\catcode `\\12\catcode `\$12\catcode
  `\&12\catcode `\#12\catcode `\^12\catcode `\_12\catcode `\%12\relax}%
\providecommand \@@startlink[1]{}%
\providecommand \@@endlink[0]{}%
\providecommand \url  [0]{\begingroup\@sanitize@url \@url }%
\providecommand \@url [1]{\endgroup\@href {#1}{\urlprefix }}%
\providecommand \urlprefix  [0]{URL }%
\providecommand \Eprint [0]{\href }%
\providecommand \doibase [0]{https://doi.org/}%
\providecommand \selectlanguage [0]{\@gobble}%
\providecommand \bibinfo  [0]{\@secondoftwo}%
\providecommand \bibfield  [0]{\@secondoftwo}%
\providecommand \translation [1]{[#1]}%
\providecommand \BibitemOpen [0]{}%
\providecommand \bibitemStop [0]{}%
\providecommand \bibitemNoStop [0]{.\EOS\space}%
\providecommand \EOS [0]{\spacefactor3000\relax}%
\providecommand \BibitemShut  [1]{\csname bibitem#1\endcsname}%
\let\auto@bib@innerbib\@empty
%</preamble>
\bibitem [{\citenamefont {Mamin}\ \emph {et~al.}(2013)\citenamefont {Mamin},
  \citenamefont {Kim}, \citenamefont {Sherwood}, \citenamefont {Rettner},
  \citenamefont {Ohno}, \citenamefont {Awschalom},\ and\ \citenamefont
  {Rugar}}]{Mamin2013a}%
  \BibitemOpen
  \bibfield  {author} {\bibinfo {author} {\bibfnamefont {H.~J.}\ \bibnamefont
  {Mamin}}, \bibinfo {author} {\bibfnamefont {M.}~\bibnamefont {Kim}}, \bibinfo
  {author} {\bibfnamefont {M.~H.}\ \bibnamefont {Sherwood}}, \bibinfo {author}
  {\bibfnamefont {C.~T.}\ \bibnamefont {Rettner}}, \bibinfo {author}
  {\bibfnamefont {K.}~\bibnamefont {Ohno}}, \bibinfo {author} {\bibfnamefont
  {D.~D.}\ \bibnamefont {Awschalom}},\ and\ \bibinfo {author} {\bibfnamefont
  {D.}~\bibnamefont {Rugar}},\ }\bibfield  {title} {\bibinfo {title} {Nanoscale
  nuclear magnetic resonance with a nitrogen-vacancy spin sensor},\ }\href
  {https://doi.org/10.1126/science.1231540} {\bibfield  {journal} {\bibinfo
  {journal} {Science}\ }\textbf {\bibinfo {volume} {339}},\ \bibinfo {pages}
  {557} (\bibinfo {year} {2013})}\BibitemShut {NoStop}%
\bibitem [{\citenamefont {Loretz}\ \emph {et~al.}(2014)\citenamefont {Loretz},
  \citenamefont {Pezzagna}, \citenamefont {Meijer},\ and\ \citenamefont
  {Degen}}]{Loretz2014a}%
  \BibitemOpen
  \bibfield  {author} {\bibinfo {author} {\bibfnamefont {M.}~\bibnamefont
  {Loretz}}, \bibinfo {author} {\bibfnamefont {S.}~\bibnamefont {Pezzagna}},
  \bibinfo {author} {\bibfnamefont {J.}~\bibnamefont {Meijer}},\ and\ \bibinfo
  {author} {\bibfnamefont {C.~L.}\ \bibnamefont {Degen}},\ }\bibfield  {title}
  {\bibinfo {title} {Nanoscale nuclear magnetic resonance with a 1.9-nm-deep
  nitrogen-vacancy sensor},\ }\href {https://doi.org/10.1063/1.4862749}
  {\bibfield  {journal} {\bibinfo  {journal} {Appl. Phys. Lett.}\ }\textbf
  {\bibinfo {volume} {104}},\ \bibinfo {pages} {033102} (\bibinfo {year}
  {2014})}\BibitemShut {NoStop}%
\bibitem [{\citenamefont {Aslam}\ \emph {et~al.}(2017)\citenamefont {Aslam},
  \citenamefont {Pfender}, \citenamefont {Neumann}, \citenamefont {Reuter},
  \citenamefont {Zappe}, \citenamefont {F{\'{a}}varo~de Oliveira},
  \citenamefont {Denisenko}, \citenamefont {Sumiya}, \citenamefont {Onoda},
  \citenamefont {Isoya},\ and\ \citenamefont {Wrachtrup}}]{Aslam2017a}%
  \BibitemOpen
  \bibfield  {author} {\bibinfo {author} {\bibfnamefont {N.}~\bibnamefont
  {Aslam}}, \bibinfo {author} {\bibfnamefont {M.}~\bibnamefont {Pfender}},
  \bibinfo {author} {\bibfnamefont {P.}~\bibnamefont {Neumann}}, \bibinfo
  {author} {\bibfnamefont {R.}~\bibnamefont {Reuter}}, \bibinfo {author}
  {\bibfnamefont {A.}~\bibnamefont {Zappe}}, \bibinfo {author} {\bibfnamefont
  {F.}~\bibnamefont {F{\'{a}}varo~de Oliveira}}, \bibinfo {author}
  {\bibfnamefont {A.}~\bibnamefont {Denisenko}}, \bibinfo {author}
  {\bibfnamefont {H.}~\bibnamefont {Sumiya}}, \bibinfo {author} {\bibfnamefont
  {S.}~\bibnamefont {Onoda}}, \bibinfo {author} {\bibfnamefont
  {J.}~\bibnamefont {Isoya}},\ and\ \bibinfo {author} {\bibfnamefont
  {J.}~\bibnamefont {Wrachtrup}},\ }\bibfield  {title} {\bibinfo {title}
  {Nanoscale nuclear magnetic resonance with chemical resolution},\ }\href
  {https://doi.org/10.1126/science.aam8697} {\bibfield  {journal} {\bibinfo
  {journal} {Science}\ }\textbf {\bibinfo {volume} {357}},\ \bibinfo {pages}
  {67} (\bibinfo {year} {2017})}\BibitemShut {NoStop}%
\bibitem [{\citenamefont {Abobeih}\ \emph {et~al.}(2019)\citenamefont
  {Abobeih}, \citenamefont {Randall}, \citenamefont {Bradley}, \citenamefont
  {Bartling}, \citenamefont {Bakker}, \citenamefont {Degen}, \citenamefont
  {Markham}, \citenamefont {Twitchen},\ and\ \citenamefont
  {Taminiau}}]{Abobeih2019b}%
  \BibitemOpen
  \bibfield  {author} {\bibinfo {author} {\bibfnamefont {M.~H.}\ \bibnamefont
  {Abobeih}}, \bibinfo {author} {\bibfnamefont {J.}~\bibnamefont {Randall}},
  \bibinfo {author} {\bibfnamefont {C.~E.}\ \bibnamefont {Bradley}}, \bibinfo
  {author} {\bibfnamefont {H.~P.}\ \bibnamefont {Bartling}}, \bibinfo {author}
  {\bibfnamefont {M.~A.}\ \bibnamefont {Bakker}}, \bibinfo {author}
  {\bibfnamefont {M.~J.}\ \bibnamefont {Degen}}, \bibinfo {author}
  {\bibfnamefont {M.}~\bibnamefont {Markham}}, \bibinfo {author} {\bibfnamefont
  {D.~J.}\ \bibnamefont {Twitchen}},\ and\ \bibinfo {author} {\bibfnamefont
  {T.~H.}\ \bibnamefont {Taminiau}},\ }\bibfield  {title} {\bibinfo {title}
  {Atomic-scale imaging of a 27-nuclear-spin cluster using a quantum sensor},\
  }\href {https://doi.org/10.1038/s41586-019-1834-7} {\bibfield  {journal}
  {\bibinfo  {journal} {Nature}\ }\textbf {\bibinfo {volume} {576}},\ \bibinfo
  {pages} {411} (\bibinfo {year} {2019})}\BibitemShut {NoStop}%
\bibitem [{\citenamefont {Smits}\ \emph {et~al.}(2019)\citenamefont {Smits},
  \citenamefont {Damron}, \citenamefont {Kehayias}, \citenamefont {McDowell},
  \citenamefont {Mosavian}, \citenamefont {Fescenko}, \citenamefont {Ristoff},
  \citenamefont {Laraoui}, \citenamefont {Jarmola},\ and\ \citenamefont
  {Acosta}}]{Smits2019a}%
  \BibitemOpen
  \bibfield  {author} {\bibinfo {author} {\bibfnamefont {J.}~\bibnamefont
  {Smits}}, \bibinfo {author} {\bibfnamefont {J.~T.}\ \bibnamefont {Damron}},
  \bibinfo {author} {\bibfnamefont {P.}~\bibnamefont {Kehayias}}, \bibinfo
  {author} {\bibfnamefont {A.~F.}\ \bibnamefont {McDowell}}, \bibinfo {author}
  {\bibfnamefont {N.}~\bibnamefont {Mosavian}}, \bibinfo {author}
  {\bibfnamefont {I.}~\bibnamefont {Fescenko}}, \bibinfo {author}
  {\bibfnamefont {N.}~\bibnamefont {Ristoff}}, \bibinfo {author} {\bibfnamefont
  {A.}~\bibnamefont {Laraoui}}, \bibinfo {author} {\bibfnamefont
  {A.}~\bibnamefont {Jarmola}},\ and\ \bibinfo {author} {\bibfnamefont {V.~M.}\
  \bibnamefont {Acosta}},\ }\bibfield  {title} {\bibinfo {title}
  {Two-dimensional nuclear magnetic resonance spectroscopy with a microfluidic
  diamond quantum sensor},\ }\href {https://doi.org/10.1126/sciadv.aaw7895}
  {\bibfield  {journal} {\bibinfo  {journal} {Sci. Adv.}\ }\textbf {\bibinfo
  {volume} {5}},\ \bibinfo {pages} {aaw7895} (\bibinfo {year}
  {2019})}\BibitemShut {NoStop}%
\bibitem [{\citenamefont {Lovchinsky}\ \emph {et~al.}(2016)\citenamefont
  {Lovchinsky}, \citenamefont {Sushkov}, \citenamefont {Urbach}, \citenamefont
  {de~Leon}, \citenamefont {Choi}, \citenamefont {De~Greve}, \citenamefont
  {Evans}, \citenamefont {Gertner}, \citenamefont {Bersin}, \citenamefont
  {M\"{u}ller}, \citenamefont {McGuinness}, \citenamefont {Jelezko},
  \citenamefont {Walsworth}, \citenamefont {Park},\ and\ \citenamefont
  {Lukin}}]{Lovchinsky2016}%
  \BibitemOpen
  \bibfield  {author} {\bibinfo {author} {\bibfnamefont {I.}~\bibnamefont
  {Lovchinsky}}, \bibinfo {author} {\bibfnamefont {A.~O.}\ \bibnamefont
  {Sushkov}}, \bibinfo {author} {\bibfnamefont {E.}~\bibnamefont {Urbach}},
  \bibinfo {author} {\bibfnamefont {N.~P.}\ \bibnamefont {de~Leon}}, \bibinfo
  {author} {\bibfnamefont {S.}~\bibnamefont {Choi}}, \bibinfo {author}
  {\bibfnamefont {K.}~\bibnamefont {De~Greve}}, \bibinfo {author}
  {\bibfnamefont {R.}~\bibnamefont {Evans}}, \bibinfo {author} {\bibfnamefont
  {R.}~\bibnamefont {Gertner}}, \bibinfo {author} {\bibfnamefont
  {E.}~\bibnamefont {Bersin}}, \bibinfo {author} {\bibfnamefont
  {C.}~\bibnamefont {M\"{u}ller}}, \bibinfo {author} {\bibfnamefont
  {L.}~\bibnamefont {McGuinness}}, \bibinfo {author} {\bibfnamefont
  {F.}~\bibnamefont {Jelezko}}, \bibinfo {author} {\bibfnamefont {R.~L.}\
  \bibnamefont {Walsworth}}, \bibinfo {author} {\bibfnamefont {H.}~\bibnamefont
  {Park}},\ and\ \bibinfo {author} {\bibfnamefont {M.~D.}\ \bibnamefont
  {Lukin}},\ }\bibfield  {title} {\bibinfo {title} {Nuclear magnetic resonance
  detection and spectroscopy of single proteins using quantum logic},\ }\href
  {https://doi.org/10.1126/science.aad8022} {\bibfield  {journal} {\bibinfo
  {journal} {Science}\ }\textbf {\bibinfo {volume} {351}},\ \bibinfo {pages}
  {836} (\bibinfo {year} {2016})}\BibitemShut {NoStop}%
\bibitem [{\citenamefont {Rosskopf}\ \emph {et~al.}(2017)\citenamefont
  {Rosskopf}, \citenamefont {Zopes}, \citenamefont {Boss},\ and\ \citenamefont
  {Degen}}]{Rosskopf2017}%
  \BibitemOpen
  \bibfield  {author} {\bibinfo {author} {\bibfnamefont {T.}~\bibnamefont
  {Rosskopf}}, \bibinfo {author} {\bibfnamefont {J.}~\bibnamefont {Zopes}},
  \bibinfo {author} {\bibfnamefont {J.~M.}\ \bibnamefont {Boss}},\ and\
  \bibinfo {author} {\bibfnamefont {C.~L.}\ \bibnamefont {Degen}},\ }\bibfield
  {title} {\bibinfo {title} {A quantum spectrum analyzer enhanced by a nuclear
  spin memory},\ }\href {https://doi.org/10.1038/s41534-017-0030-6} {\bibfield
  {journal} {\bibinfo  {journal} {npj Quantum Inf.}\ }\textbf {\bibinfo
  {volume} {3}},\ \bibinfo {pages} {33} (\bibinfo {year} {2017})}\BibitemShut
  {NoStop}%
\bibitem [{\citenamefont {Smeltzer}\ \emph {et~al.}(2009)\citenamefont
  {Smeltzer}, \citenamefont {McIntyre},\ and\ \citenamefont
  {Childress}}]{Smeltzer2009}%
  \BibitemOpen
  \bibfield  {author} {\bibinfo {author} {\bibfnamefont {B.}~\bibnamefont
  {Smeltzer}}, \bibinfo {author} {\bibfnamefont {J.}~\bibnamefont {McIntyre}},\
  and\ \bibinfo {author} {\bibfnamefont {L.}~\bibnamefont {Childress}},\
  }\bibfield  {title} {\bibinfo {title} {Robust control of individual nuclear
  spins in diamond},\ }\href {https://doi.org/10.1103/PhysRevA.80.050302}
  {\bibfield  {journal} {\bibinfo  {journal} {Phys. Rev. A}\ }\textbf {\bibinfo
  {volume} {80}},\ \bibinfo {pages} {050302} (\bibinfo {year}
  {2009})}\BibitemShut {NoStop}%
\bibitem [{\citenamefont {Degen}\ \emph {et~al.}(2017)\citenamefont {Degen},
  \citenamefont {Reinhard},\ and\ \citenamefont {Cappellaro}}]{Degen2017}%
  \BibitemOpen
  \bibfield  {author} {\bibinfo {author} {\bibfnamefont {C.~L.}\ \bibnamefont
  {Degen}}, \bibinfo {author} {\bibfnamefont {F.}~\bibnamefont {Reinhard}},\
  and\ \bibinfo {author} {\bibfnamefont {P.}~\bibnamefont {Cappellaro}},\
  }\bibfield  {title} {\bibinfo {title} {Quantum sensing},\ }\href
  {https://doi.org/10.1103/RevModPhys.89.035002} {\bibfield  {journal}
  {\bibinfo  {journal} {Rev. Mod. Phys.}\ }\textbf {\bibinfo {volume} {89}},\
  \bibinfo {pages} {035002} (\bibinfo {year} {2017})}\BibitemShut {NoStop}%
\bibitem [{\citenamefont {Laraoui}\ \emph {et~al.}(2013)\citenamefont
  {Laraoui}, \citenamefont {Dolde}, \citenamefont {Burk}, \citenamefont
  {Reinhard}, \citenamefont {Wrachtrup},\ and\ \citenamefont
  {Meriles}}]{Laraoui2013a}%
  \BibitemOpen
  \bibfield  {author} {\bibinfo {author} {\bibfnamefont {A.}~\bibnamefont
  {Laraoui}}, \bibinfo {author} {\bibfnamefont {F.}~\bibnamefont {Dolde}},
  \bibinfo {author} {\bibfnamefont {C.}~\bibnamefont {Burk}}, \bibinfo {author}
  {\bibfnamefont {F.}~\bibnamefont {Reinhard}}, \bibinfo {author}
  {\bibfnamefont {J.}~\bibnamefont {Wrachtrup}},\ and\ \bibinfo {author}
  {\bibfnamefont {C.~A.}\ \bibnamefont {Meriles}},\ }\bibfield  {title}
  {\bibinfo {title} {High-resolution correlation spectroscopy of $^{13}$c spins
  near a nitrogen-vacancy centre in diamond},\ }\href
  {https://doi.org/10.1038/ncomms2685} {\bibfield  {journal} {\bibinfo
  {journal} {Nat. Commun.}\ }\textbf {\bibinfo {volume} {4}},\ \bibinfo {pages}
  {1651} (\bibinfo {year} {2013})}\BibitemShut {NoStop}%
\bibitem [{\citenamefont {Munuera-Javaloy}\ \emph {et~al.}(2023)\citenamefont
  {Munuera-Javaloy}, \citenamefont {Tobalina},\ and\ \citenamefont
  {Casanova}}]{Munuera2023}%
  \BibitemOpen
  \bibfield  {author} {\bibinfo {author} {\bibfnamefont {C.}~\bibnamefont
  {Munuera-Javaloy}}, \bibinfo {author} {\bibfnamefont {A.}~\bibnamefont
  {Tobalina}},\ and\ \bibinfo {author} {\bibfnamefont {J.}~\bibnamefont
  {Casanova}},\ }\bibfield  {title} {\bibinfo {title} {High-resolution {NMR}
  spectroscopy at large fields with nitrogen vacancy centers},\ }\href
  {https://doi.org/10.1103/PhysRevLett.130.133603} {\bibfield  {journal}
  {\bibinfo  {journal} {Phys. Rev. Lett.}\ }\textbf {\bibinfo {volume} {130}},\
  \bibinfo {pages} {133603} (\bibinfo {year} {2023})}\BibitemShut {NoStop}%
\bibitem [{\citenamefont {Fuchs}\ \emph {et~al.}(2009)\citenamefont {Fuchs},
  \citenamefont {Dobrovitski}, \citenamefont {Toyli}, \citenamefont
  {Heremans},\ and\ \citenamefont {Awschalom}}]{Fuchs2009a}%
  \BibitemOpen
  \bibfield  {author} {\bibinfo {author} {\bibfnamefont {G.~D.}\ \bibnamefont
  {Fuchs}}, \bibinfo {author} {\bibfnamefont {V.~V.}\ \bibnamefont
  {Dobrovitski}}, \bibinfo {author} {\bibfnamefont {D.~M.}\ \bibnamefont
  {Toyli}}, \bibinfo {author} {\bibfnamefont {F.~J.}\ \bibnamefont
  {Heremans}},\ and\ \bibinfo {author} {\bibfnamefont {D.~D.}\ \bibnamefont
  {Awschalom}},\ }\bibfield  {title} {\bibinfo {title} {Gigahertz dynamics of a
  strongly driven single quantum spin},\ }\href
  {https://doi.org/10.1126/science.1181193} {\bibfield  {journal} {\bibinfo
  {journal} {Science}\ }\textbf {\bibinfo {volume} {326}},\ \bibinfo {pages}
  {1520} (\bibinfo {year} {2009})}\BibitemShut {NoStop}%
\bibitem [{\citenamefont {Herb}\ \emph {et~al.}(2020)\citenamefont {Herb},
  \citenamefont {Zopes}, \citenamefont {Cujia},\ and\ \citenamefont
  {Degen}}]{Herb2020}%
  \BibitemOpen
  \bibfield  {author} {\bibinfo {author} {\bibfnamefont {K.}~\bibnamefont
  {Herb}}, \bibinfo {author} {\bibfnamefont {J.}~\bibnamefont {Zopes}},
  \bibinfo {author} {\bibfnamefont {K.~S.}\ \bibnamefont {Cujia}},\ and\
  \bibinfo {author} {\bibfnamefont {C.~L.}\ \bibnamefont {Degen}},\ }\bibfield
  {title} {\bibinfo {title} {Broadband radio-frequency transmitter for fast
  nuclear spin control},\ }\href {https://doi.org/10.1063/5.0013776} {\bibfield
   {journal} {\bibinfo  {journal} {Rev. Sci. Instrum.}\ }\textbf {\bibinfo
  {volume} {91}},\ \bibinfo {pages} {113106} (\bibinfo {year}
  {2020})}\BibitemShut {NoStop}%
\bibitem [{\citenamefont {Song}\ \emph {et~al.}(2016)\citenamefont {Song},
  \citenamefont {Kestner}, \citenamefont {Wang},\ and\ \citenamefont
  {Das~Sarma}}]{Song2016FastApproximation}%
  \BibitemOpen
  \bibfield  {author} {\bibinfo {author} {\bibfnamefont {Y.}~\bibnamefont
  {Song}}, \bibinfo {author} {\bibfnamefont {J.~P.}\ \bibnamefont {Kestner}},
  \bibinfo {author} {\bibfnamefont {X.}~\bibnamefont {Wang}},\ and\ \bibinfo
  {author} {\bibfnamefont {S.}~\bibnamefont {Das~Sarma}},\ }\bibfield  {title}
  {\bibinfo {title} {Fast control of semiconductor qubits beyond the
  rotating-wave approximation},\ }\href
  {https://doi.org/10.1103/PhysRevA.94.012321} {\bibfield  {journal} {\bibinfo
  {journal} {Phys. Rev. A}\ }\textbf {\bibinfo {volume} {94}},\ \bibinfo
  {pages} {012321} (\bibinfo {year} {2016})}\BibitemShut {NoStop}%
\bibitem [{\citenamefont {Yang}\ \emph {et~al.}(2017)\citenamefont {Yang},
  \citenamefont {Coppersmith},\ and\ \citenamefont
  {Friesen}}]{Yang2017AchievingQubit}%
  \BibitemOpen
  \bibfield  {author} {\bibinfo {author} {\bibfnamefont {Y.~C.}\ \bibnamefont
  {Yang}}, \bibinfo {author} {\bibfnamefont {S.~N.}\ \bibnamefont
  {Coppersmith}},\ and\ \bibinfo {author} {\bibfnamefont {M.}~\bibnamefont
  {Friesen}},\ }\bibfield  {title} {\bibinfo {title} {Achieving high-fidelity
  single-qubit gates in a strongly driven silicon-quantum-dot hybrid qubit},\
  }\href {https://doi.org/10.1103/PhysRevA.95.062321} {\bibfield  {journal}
  {\bibinfo  {journal} {Phys. Rev. A}\ }\textbf {\bibinfo {volume} {95}},\
  \bibinfo {pages} {062321} (\bibinfo {year} {2017})}\BibitemShut {NoStop}%
\bibitem [{\citenamefont {Khaneja}\ \emph {et~al.}(2001)\citenamefont
  {Khaneja}, \citenamefont {Brockett},\ and\ \citenamefont
  {Glaser}}]{Khaneja2001TimeSystems}%
  \BibitemOpen
  \bibfield  {author} {\bibinfo {author} {\bibfnamefont {N.}~\bibnamefont
  {Khaneja}}, \bibinfo {author} {\bibfnamefont {R.}~\bibnamefont {Brockett}},\
  and\ \bibinfo {author} {\bibfnamefont {S.~J.}\ \bibnamefont {Glaser}},\
  }\bibfield  {title} {\bibinfo {title} {Time optimal control in spin
  systems},\ }\href {https://doi.org/10.1103/PhysRevA.63.032308} {\bibfield
  {journal} {\bibinfo  {journal} {Phys. Rev. A}\ }\textbf {\bibinfo {volume}
  {63}},\ \bibinfo {pages} {032308} (\bibinfo {year} {2001})}\BibitemShut
  {NoStop}%
\bibitem [{\citenamefont {Boscain}\ and\ \citenamefont
  {Mason}(2006)}]{Boscain2006TimeField}%
  \BibitemOpen
  \bibfield  {author} {\bibinfo {author} {\bibfnamefont {U.}~\bibnamefont
  {Boscain}}\ and\ \bibinfo {author} {\bibfnamefont {P.}~\bibnamefont
  {Mason}},\ }\bibfield  {title} {\bibinfo {title} {Time minimal trajectories
  for a spin 1/2 particle in a magnetic field},\ }\href
  {https://doi.org/10.1063/1.2203236} {\bibfield  {journal} {\bibinfo
  {journal} {J. Math. Phys.}\ }\textbf {\bibinfo {volume} {47}},\ \bibinfo
  {pages} {062101} (\bibinfo {year} {2006})}\BibitemShut {NoStop}%
\bibitem [{\citenamefont {Wang}\ and\ \citenamefont
  {Dobrovitski}(2011)}]{Wang2011Time-optimalDiamond}%
  \BibitemOpen
  \bibfield  {author} {\bibinfo {author} {\bibfnamefont {Z.~H.}\ \bibnamefont
  {Wang}}\ and\ \bibinfo {author} {\bibfnamefont {V.~V.}\ \bibnamefont
  {Dobrovitski}},\ }\bibfield  {title} {\bibinfo {title} {Time-optimal rotation
  of a spin 1/2: {A}pplication to the {NV} center spin in diamond},\ }\href
  {https://doi.org/10.1103/PhysRevB.84.045303} {\bibfield  {journal} {\bibinfo
  {journal} {Phys. Rev. B}\ }\textbf {\bibinfo {volume} {84}},\ \bibinfo
  {pages} {045303} (\bibinfo {year} {2011})}\BibitemShut {NoStop}%
\bibitem [{\citenamefont {Avinadav}\ \emph {et~al.}(2014)\citenamefont
  {Avinadav}, \citenamefont {Fischer}, \citenamefont {London},\ and\
  \citenamefont {Gershoni}}]{Avinadav2014Time-optimalDriving}%
  \BibitemOpen
  \bibfield  {author} {\bibinfo {author} {\bibfnamefont {C.}~\bibnamefont
  {Avinadav}}, \bibinfo {author} {\bibfnamefont {R.}~\bibnamefont {Fischer}},
  \bibinfo {author} {\bibfnamefont {P.}~\bibnamefont {London}},\ and\ \bibinfo
  {author} {\bibfnamefont {D.}~\bibnamefont {Gershoni}},\ }\bibfield  {title}
  {\bibinfo {title} {Time-optimal universal control of two-level systems under
  strong driving},\ }\href {https://doi.org/10.1103/PhysRevB.89.245311}
  {\bibfield  {journal} {\bibinfo  {journal} {Phys. Rev. B}\ }\textbf {\bibinfo
  {volume} {89}},\ \bibinfo {pages} {245311} (\bibinfo {year}
  {2014})}\BibitemShut {NoStop}%
\bibitem [{\citenamefont {Scheuer}\ \emph {et~al.}(2014)\citenamefont
  {Scheuer}, \citenamefont {Kong}, \citenamefont {Said}, \citenamefont {Chen},
  \citenamefont {Kurz}, \citenamefont {Marseglia}, \citenamefont {Du},
  \citenamefont {Hemmer}, \citenamefont {Montangero}, \citenamefont {Calarco},
  \citenamefont {Naydenov},\ and\ \citenamefont
  {Jelezko}}]{Scheuer2014PreciseApproximation}%
  \BibitemOpen
  \bibfield  {author} {\bibinfo {author} {\bibfnamefont {J.}~\bibnamefont
  {Scheuer}}, \bibinfo {author} {\bibfnamefont {X.}~\bibnamefont {Kong}},
  \bibinfo {author} {\bibfnamefont {R.~S.}\ \bibnamefont {Said}}, \bibinfo
  {author} {\bibfnamefont {J.}~\bibnamefont {Chen}}, \bibinfo {author}
  {\bibfnamefont {A.}~\bibnamefont {Kurz}}, \bibinfo {author} {\bibfnamefont
  {L.}~\bibnamefont {Marseglia}}, \bibinfo {author} {\bibfnamefont
  {J.}~\bibnamefont {Du}}, \bibinfo {author} {\bibfnamefont {P.~R.}\
  \bibnamefont {Hemmer}}, \bibinfo {author} {\bibfnamefont {S.}~\bibnamefont
  {Montangero}}, \bibinfo {author} {\bibfnamefont {T.}~\bibnamefont {Calarco}},
  \bibinfo {author} {\bibfnamefont {B.}~\bibnamefont {Naydenov}},\ and\
  \bibinfo {author} {\bibfnamefont {F.}~\bibnamefont {Jelezko}},\ }\bibfield
  {title} {\bibinfo {title} {Precise qubit control beyond the rotating wave
  approximation},\ }\href {https://doi.org/10.1088/1367-2630/16/9/093022}
  {\bibfield  {journal} {\bibinfo  {journal} {New J. Phys.}\ }\textbf {\bibinfo
  {volume} {16}},\ \bibinfo {pages} {093022} (\bibinfo {year}
  {2014})}\BibitemShut {NoStop}%
\bibitem [{\citenamefont {Degen}\ \emph {et~al.}(2009)\citenamefont {Degen},
  \citenamefont {Poggio}, \citenamefont {Mamin}, \citenamefont {Rettner},\ and\
  \citenamefont {Rugar}}]{Degen2009a}%
  \BibitemOpen
  \bibfield  {author} {\bibinfo {author} {\bibfnamefont {C.~L.}\ \bibnamefont
  {Degen}}, \bibinfo {author} {\bibfnamefont {M.}~\bibnamefont {Poggio}},
  \bibinfo {author} {\bibfnamefont {H.~J.}\ \bibnamefont {Mamin}}, \bibinfo
  {author} {\bibfnamefont {C.~T.}\ \bibnamefont {Rettner}},\ and\ \bibinfo
  {author} {\bibfnamefont {D.}~\bibnamefont {Rugar}},\ }\bibfield  {title}
  {\bibinfo {title} {Nanoscale magnetic resonance imaging},\ }\href
  {https://doi.org/10.1073/pnas.0812068106} {\bibfield  {journal} {\bibinfo
  {journal} {Proc. Natl. Acad. Sci.}\ }\textbf {\bibinfo {volume} {106}},\
  \bibinfo {pages} {1313} (\bibinfo {year} {2009})}\BibitemShut {NoStop}%
\bibitem [{\citenamefont {Grinolds}\ \emph {et~al.}(2014)\citenamefont
  {Grinolds}, \citenamefont {Warner}, \citenamefont {De~Greve}, \citenamefont
  {Dovzhenko}, \citenamefont {Thiel}, \citenamefont {Walsworth}, \citenamefont
  {Hong}, \citenamefont {Maletinsky},\ and\ \citenamefont
  {Yacoby}}]{Grinolds2014}%
  \BibitemOpen
  \bibfield  {author} {\bibinfo {author} {\bibfnamefont {M.~S.}\ \bibnamefont
  {Grinolds}}, \bibinfo {author} {\bibfnamefont {M.}~\bibnamefont {Warner}},
  \bibinfo {author} {\bibfnamefont {K.}~\bibnamefont {De~Greve}}, \bibinfo
  {author} {\bibfnamefont {Y.}~\bibnamefont {Dovzhenko}}, \bibinfo {author}
  {\bibfnamefont {L.}~\bibnamefont {Thiel}}, \bibinfo {author} {\bibfnamefont
  {R.~L.}\ \bibnamefont {Walsworth}}, \bibinfo {author} {\bibfnamefont
  {S.}~\bibnamefont {Hong}}, \bibinfo {author} {\bibfnamefont {P.}~\bibnamefont
  {Maletinsky}},\ and\ \bibinfo {author} {\bibfnamefont {A.}~\bibnamefont
  {Yacoby}},\ }\bibfield  {title} {\bibinfo {title} {{Subnanometre resolution
  in three-dimensional magnetic resonance imaging of individual dark spins}},\
  }\href {https://doi.org/10.1038/nnano.2014.30} {\bibfield  {journal}
  {\bibinfo  {journal} {Nat. Nanotechnol.}\ }\textbf {\bibinfo {volume} {9}},\
  \bibinfo {pages} {279} (\bibinfo {year} {2014})}\BibitemShut {NoStop}%
\bibitem [{\citenamefont {Neumann}\ \emph {et~al.}(2010)\citenamefont
  {Neumann}, \citenamefont {Beck}, \citenamefont {Steiner}, \citenamefont
  {Rempp}, \citenamefont {Fedder}, \citenamefont {Hemmer}, \citenamefont
  {Wrachtrup},\ and\ \citenamefont {Jelezko}}]{Neumann2010}%
  \BibitemOpen
  \bibfield  {author} {\bibinfo {author} {\bibfnamefont {P.}~\bibnamefont
  {Neumann}}, \bibinfo {author} {\bibfnamefont {J.}~\bibnamefont {Beck}},
  \bibinfo {author} {\bibfnamefont {M.}~\bibnamefont {Steiner}}, \bibinfo
  {author} {\bibfnamefont {F.}~\bibnamefont {Rempp}}, \bibinfo {author}
  {\bibfnamefont {H.}~\bibnamefont {Fedder}}, \bibinfo {author} {\bibfnamefont
  {P.~R.}\ \bibnamefont {Hemmer}}, \bibinfo {author} {\bibfnamefont
  {J.}~\bibnamefont {Wrachtrup}},\ and\ \bibinfo {author} {\bibfnamefont
  {F.}~\bibnamefont {Jelezko}},\ }\bibfield  {title} {\bibinfo {title}
  {Single-shot readout of a single nuclear spin},\ }\href
  {https://doi.org/10.1126/science.1189075} {\bibfield  {journal} {\bibinfo
  {journal} {Science}\ }\textbf {\bibinfo {volume} {329}},\ \bibinfo {pages}
  {542} (\bibinfo {year} {2010})}\BibitemShut {NoStop}%
\bibitem [{\citenamefont {Staudacher}\ \emph {et~al.}(2013)\citenamefont
  {Staudacher}, \citenamefont {Shi}, \citenamefont {Pezzagna}, \citenamefont
  {Meijer}, \citenamefont {Du}, \citenamefont {Meriles}, \citenamefont
  {Reinhard},\ and\ \citenamefont {Wrachtrup}}]{Staudacher2013}%
  \BibitemOpen
  \bibfield  {author} {\bibinfo {author} {\bibfnamefont {T.}~\bibnamefont
  {Staudacher}}, \bibinfo {author} {\bibfnamefont {F.}~\bibnamefont {Shi}},
  \bibinfo {author} {\bibfnamefont {S.}~\bibnamefont {Pezzagna}}, \bibinfo
  {author} {\bibfnamefont {J.}~\bibnamefont {Meijer}}, \bibinfo {author}
  {\bibfnamefont {J.}~\bibnamefont {Du}}, \bibinfo {author} {\bibfnamefont
  {C.~A.}\ \bibnamefont {Meriles}}, \bibinfo {author} {\bibfnamefont
  {F.}~\bibnamefont {Reinhard}},\ and\ \bibinfo {author} {\bibfnamefont
  {J.}~\bibnamefont {Wrachtrup}},\ }\bibfield  {title} {\bibinfo {title}
  {Nuclear magnetic resonance spectroscopy on a (5-nanometer)$^3$ sample
  volume},\ }\href {https://doi.org/10.1126/science.1231675} {\bibfield
  {journal} {\bibinfo  {journal} {Science}\ }\textbf {\bibinfo {volume}
  {339}},\ \bibinfo {pages} {561} (\bibinfo {year} {2013})}\BibitemShut
  {NoStop}%
\bibitem [{\citenamefont {Wang}\ \emph {et~al.}(2019)\citenamefont {Wang},
  \citenamefont {Lang}, \citenamefont {Schmitt}, \citenamefont {Lang},
  \citenamefont {Casanova}, \citenamefont {McGuinness}, \citenamefont
  {Monteiro}, \citenamefont {Jelezko},\ and\ \citenamefont
  {Plenio}}]{Wang2019a}%
  \BibitemOpen
  \bibfield  {author} {\bibinfo {author} {\bibfnamefont {Z.-Y.}\ \bibnamefont
  {Wang}}, \bibinfo {author} {\bibfnamefont {J.~E.}\ \bibnamefont {Lang}},
  \bibinfo {author} {\bibfnamefont {S.}~\bibnamefont {Schmitt}}, \bibinfo
  {author} {\bibfnamefont {J.}~\bibnamefont {Lang}}, \bibinfo {author}
  {\bibfnamefont {J.}~\bibnamefont {Casanova}}, \bibinfo {author}
  {\bibfnamefont {L.}~\bibnamefont {McGuinness}}, \bibinfo {author}
  {\bibfnamefont {T.~S.}\ \bibnamefont {Monteiro}}, \bibinfo {author}
  {\bibfnamefont {F.}~\bibnamefont {Jelezko}},\ and\ \bibinfo {author}
  {\bibfnamefont {M.~B.}\ \bibnamefont {Plenio}},\ }\bibfield  {title}
  {\bibinfo {title} {Randomization of pulse phases for unambiguous and robust
  quantum sensing},\ }\href {https://doi.org/10.1103/PhysRevLett.122.200403}
  {\bibfield  {journal} {\bibinfo  {journal} {Phys. Rev. Lett.}\ }\textbf
  {\bibinfo {volume} {122}},\ \bibinfo {pages} {200403} (\bibinfo {year}
  {2019})}\BibitemShut {NoStop}%
\bibitem [{\citenamefont {Pfender}\ \emph {et~al.}(2017)\citenamefont
  {Pfender}, \citenamefont {Aslam}, \citenamefont {Sumiya}, \citenamefont
  {Onoda}, \citenamefont {Neumann}, \citenamefont {Isoya}, \citenamefont
  {Meriles},\ and\ \citenamefont {Wrachtrup}}]{Pfender2017a}%
  \BibitemOpen
  \bibfield  {author} {\bibinfo {author} {\bibfnamefont {M.}~\bibnamefont
  {Pfender}}, \bibinfo {author} {\bibfnamefont {N.}~\bibnamefont {Aslam}},
  \bibinfo {author} {\bibfnamefont {H.}~\bibnamefont {Sumiya}}, \bibinfo
  {author} {\bibfnamefont {S.}~\bibnamefont {Onoda}}, \bibinfo {author}
  {\bibfnamefont {P.}~\bibnamefont {Neumann}}, \bibinfo {author} {\bibfnamefont
  {J.}~\bibnamefont {Isoya}}, \bibinfo {author} {\bibfnamefont {C.~A.}\
  \bibnamefont {Meriles}},\ and\ \bibinfo {author} {\bibfnamefont
  {J.}~\bibnamefont {Wrachtrup}},\ }\bibfield  {title} {\bibinfo {title}
  {{Nonvolatile nuclear spin memory enables sensor-unlimited nanoscale
  spectroscopy of small spin clusters}},\ }\href
  {https://doi.org/10.1038/s41467-017-00964-z} {\bibfield  {journal} {\bibinfo
  {journal} {Nat. Commun.}\ }\textbf {\bibinfo {volume} {8}},\ \bibinfo {pages}
  {834} (\bibinfo {year} {2017})}\BibitemShut {NoStop}%
\bibitem [{\citenamefont {Laraoui}\ \emph {et~al.}(2015)\citenamefont
  {Laraoui}, \citenamefont {Pagliero},\ and\ \citenamefont
  {Meriles}}]{Laraoui2015a}%
  \BibitemOpen
  \bibfield  {author} {\bibinfo {author} {\bibfnamefont {A.}~\bibnamefont
  {Laraoui}}, \bibinfo {author} {\bibfnamefont {D.}~\bibnamefont {Pagliero}},\
  and\ \bibinfo {author} {\bibfnamefont {C.~A.}\ \bibnamefont {Meriles}},\
  }\bibfield  {title} {\bibinfo {title} {Imaging nuclear spins weakly coupled
  to a probe paramagnetic center},\ }\href
  {https://doi.org/10.1103/PhysRevB.91.205410} {\bibfield  {journal} {\bibinfo
  {journal} {Phys. Rev. B}\ }\textbf {\bibinfo {volume} {91}},\ \bibinfo
  {pages} {205410} (\bibinfo {year} {2015})}\BibitemShut {NoStop}%
\bibitem [{\citenamefont {Staudacher}\ \emph {et~al.}(2015)\citenamefont
  {Staudacher}, \citenamefont {Raatz}, \citenamefont {Pezzagna}, \citenamefont
  {Meijer}, \citenamefont {Reinhard}, \citenamefont {Meriles},\ and\
  \citenamefont {Wrachtrup}}]{Staudacher2015a}%
  \BibitemOpen
  \bibfield  {author} {\bibinfo {author} {\bibfnamefont {T.}~\bibnamefont
  {Staudacher}}, \bibinfo {author} {\bibfnamefont {N.}~\bibnamefont {Raatz}},
  \bibinfo {author} {\bibfnamefont {S.}~\bibnamefont {Pezzagna}}, \bibinfo
  {author} {\bibfnamefont {J.}~\bibnamefont {Meijer}}, \bibinfo {author}
  {\bibfnamefont {F.}~\bibnamefont {Reinhard}}, \bibinfo {author}
  {\bibfnamefont {C.~A.}\ \bibnamefont {Meriles}},\ and\ \bibinfo {author}
  {\bibfnamefont {J.}~\bibnamefont {Wrachtrup}},\ }\bibfield  {title} {\bibinfo
  {title} {Probing molecular dynamics at the nanoscale via an individual
  paramagnetic centre},\ }\href {https://doi.org/10.1038/ncomms9527} {\bibfield
   {journal} {\bibinfo  {journal} {Nat. Commun.}\ }\textbf {\bibinfo {volume}
  {6}},\ \bibinfo {pages} {8527} (\bibinfo {year} {2015})}\BibitemShut
  {NoStop}%
\bibitem [{\citenamefont {Nguyen}\ \emph {et~al.}(2019)\citenamefont {Nguyen},
  \citenamefont {Sukachev}, \citenamefont {Bhaskar}, \citenamefont {Machielse},
  \citenamefont {Levonian}, \citenamefont {Knall}, \citenamefont {Stroganov},
  \citenamefont {Chia}, \citenamefont {Burek}, \citenamefont {Riedinger},
  \citenamefont {Park}, \citenamefont {Lončar},\ and\ \citenamefont
  {Lukin}}]{Nguyen2019}%
  \BibitemOpen
  \bibfield  {author} {\bibinfo {author} {\bibfnamefont {C.~T.}\ \bibnamefont
  {Nguyen}}, \bibinfo {author} {\bibfnamefont {D.~D.}\ \bibnamefont
  {Sukachev}}, \bibinfo {author} {\bibfnamefont {M.~K.}\ \bibnamefont
  {Bhaskar}}, \bibinfo {author} {\bibfnamefont {B.}~\bibnamefont {Machielse}},
  \bibinfo {author} {\bibfnamefont {D.~S.}\ \bibnamefont {Levonian}}, \bibinfo
  {author} {\bibfnamefont {E.~N.}\ \bibnamefont {Knall}}, \bibinfo {author}
  {\bibfnamefont {P.}~\bibnamefont {Stroganov}}, \bibinfo {author}
  {\bibfnamefont {C.}~\bibnamefont {Chia}}, \bibinfo {author} {\bibfnamefont
  {M.~J.}\ \bibnamefont {Burek}}, \bibinfo {author} {\bibfnamefont
  {R.}~\bibnamefont {Riedinger}}, \bibinfo {author} {\bibfnamefont
  {H.}~\bibnamefont {Park}}, \bibinfo {author} {\bibfnamefont {M.}~\bibnamefont
  {Lončar}},\ and\ \bibinfo {author} {\bibfnamefont {M.~D.}\ \bibnamefont
  {Lukin}},\ }\bibfield  {title} {\bibinfo {title} {An integrated nanophotonic
  quantum register based on silicon-vacancy spins in diamond},\ }\href
  {https://doi.org/10.1103/PhysRevB.100.165428} {\bibfield  {journal} {\bibinfo
   {journal} {Phys. Rev. B}\ }\textbf {\bibinfo {volume} {100}},\ \bibinfo
  {pages} {165428} (\bibinfo {year} {2019})}\BibitemShut {NoStop}%
\bibitem [{\citenamefont {Miao}\ \emph {et~al.}(2020)\citenamefont {Miao},
  \citenamefont {Blanton}, \citenamefont {Anderson}, \citenamefont {Bourassa},
  \citenamefont {Crook}, \citenamefont {Wolfowicz}, \citenamefont {Abe},
  \citenamefont {Ohshima},\ and\ \citenamefont
  {Awschalom}}]{miao2020universal}%
  \BibitemOpen
  \bibfield  {author} {\bibinfo {author} {\bibfnamefont {K.~C.}\ \bibnamefont
  {Miao}}, \bibinfo {author} {\bibfnamefont {J.~P.}\ \bibnamefont {Blanton}},
  \bibinfo {author} {\bibfnamefont {C.~P.}\ \bibnamefont {Anderson}}, \bibinfo
  {author} {\bibfnamefont {A.}~\bibnamefont {Bourassa}}, \bibinfo {author}
  {\bibfnamefont {A.~L.}\ \bibnamefont {Crook}}, \bibinfo {author}
  {\bibfnamefont {G.}~\bibnamefont {Wolfowicz}}, \bibinfo {author}
  {\bibfnamefont {H.}~\bibnamefont {Abe}}, \bibinfo {author} {\bibfnamefont
  {T.}~\bibnamefont {Ohshima}},\ and\ \bibinfo {author} {\bibfnamefont {D.~D.}\
  \bibnamefont {Awschalom}},\ }\bibfield  {title} {\bibinfo {title} {Universal
  coherence protection in a solid-state spin qubit},\ }\href@noop {} {\bibfield
   {journal} {\bibinfo  {journal} {Science}\ }\textbf {\bibinfo {volume}
  {369}},\ \bibinfo {pages} {1493} (\bibinfo {year} {2020})}\BibitemShut
  {NoStop}%
\bibitem [{\citenamefont {Stern}\ \emph {et~al.}(2022)\citenamefont {Stern},
  \citenamefont {Gu}, \citenamefont {Jarman}, \citenamefont {Eizagirre~Barker},
  \citenamefont {Mendelson}, \citenamefont {Chugh}, \citenamefont {Schott},
  \citenamefont {Tan}, \citenamefont {Sirringhaus}, \citenamefont
  {Aharonovich},\ and\ \citenamefont
  {Atat{\"{u}}re}}]{Stern2022Room-temperatureNitride}%
  \BibitemOpen
  \bibfield  {author} {\bibinfo {author} {\bibfnamefont {H.~L.}\ \bibnamefont
  {Stern}}, \bibinfo {author} {\bibfnamefont {Q.}~\bibnamefont {Gu}}, \bibinfo
  {author} {\bibfnamefont {J.}~\bibnamefont {Jarman}}, \bibinfo {author}
  {\bibfnamefont {S.}~\bibnamefont {Eizagirre~Barker}}, \bibinfo {author}
  {\bibfnamefont {N.}~\bibnamefont {Mendelson}}, \bibinfo {author}
  {\bibfnamefont {D.}~\bibnamefont {Chugh}}, \bibinfo {author} {\bibfnamefont
  {S.}~\bibnamefont {Schott}}, \bibinfo {author} {\bibfnamefont {H.~H.}\
  \bibnamefont {Tan}}, \bibinfo {author} {\bibfnamefont {H.}~\bibnamefont
  {Sirringhaus}}, \bibinfo {author} {\bibfnamefont {I.}~\bibnamefont
  {Aharonovich}},\ and\ \bibinfo {author} {\bibfnamefont {M.}~\bibnamefont
  {Atat{\"{u}}re}},\ }\bibfield  {title} {\bibinfo {title} {Room-temperature
  optically detected magnetic resonance of single defects in hexagonal boron
  nitride},\ }\href {https://doi.org/10.1038/s41467-022-28169-z} {\bibfield
  {journal} {\bibinfo  {journal} {Nat. Commun.}\ }\textbf {\bibinfo {volume}
  {13}},\ \bibinfo {pages} {618} (\bibinfo {year} {2022})}\BibitemShut
  {NoStop}%
\bibitem [{\citenamefont {Yan}\ \emph {et~al.}(2013)\citenamefont {Yan},
  \citenamefont {Gustavsson}, \citenamefont {Bylander}, \citenamefont {Jin},
  \citenamefont {Yoshihara}, \citenamefont {Cory}, \citenamefont {Nakamura},
  \citenamefont {Orlando},\ and\ \citenamefont
  {Oliver}}]{Yan2013Rotating-frameEvolution}%
  \BibitemOpen
  \bibfield  {author} {\bibinfo {author} {\bibfnamefont {F.}~\bibnamefont
  {Yan}}, \bibinfo {author} {\bibfnamefont {S.}~\bibnamefont {Gustavsson}},
  \bibinfo {author} {\bibfnamefont {J.}~\bibnamefont {Bylander}}, \bibinfo
  {author} {\bibfnamefont {X.}~\bibnamefont {Jin}}, \bibinfo {author}
  {\bibfnamefont {F.}~\bibnamefont {Yoshihara}}, \bibinfo {author}
  {\bibfnamefont {D.~G.}\ \bibnamefont {Cory}}, \bibinfo {author}
  {\bibfnamefont {Y.}~\bibnamefont {Nakamura}}, \bibinfo {author}
  {\bibfnamefont {T.~P.}\ \bibnamefont {Orlando}},\ and\ \bibinfo {author}
  {\bibfnamefont {W.~D.}\ \bibnamefont {Oliver}},\ }\bibfield  {title}
  {\bibinfo {title} {Rotating-frame relaxation as a noise spectrum analyser of
  a superconducting qubit undergoing driven evolution},\ }\href
  {https://doi.org/10.1038/ncomms3337} {\bibfield  {journal} {\bibinfo
  {journal} {Nat. Commun.}\ }\textbf {\bibinfo {volume} {4}},\ \bibinfo {pages}
  {2337} (\bibinfo {year} {2013})}\BibitemShut {NoStop}%
\bibitem [{\citenamefont {Slichter}(1990)}]{Slichter1990}%
  \BibitemOpen
  \bibfield  {author} {\bibinfo {author} {\bibfnamefont {C.~P.}\ \bibnamefont
  {Slichter}},\ }\href {https://doi.org/10.1007/978-3-662-09441-9} {\emph
  {\bibinfo {title} {Principles of Magnetic Resonance}}},\ Vol.~\bibinfo
  {volume} {1}\ (\bibinfo  {publisher} {Springer Berlin Heidelberg},\ \bibinfo
  {year} {1990})\BibitemShut {NoStop}%
\bibitem [{\citenamefont
  {MacLaughlin}(1970)}]{MacLaughlin1970SuppressionResonance}%
  \BibitemOpen
  \bibfield  {author} {\bibinfo {author} {\bibfnamefont {D.~E.}\ \bibnamefont
  {MacLaughlin}},\ }\bibfield  {title} {\bibinfo {title} {Suppression of
  nutation angle variation in pulsed nuclear magnetic resonance},\ }\href
  {https://doi.org/10.1063/1.1684760} {\bibfield  {journal} {\bibinfo
  {journal} {Rev. Sci. Instrum.}\ }\textbf {\bibinfo {volume} {41}},\ \bibinfo
  {pages} {1202} (\bibinfo {year} {1970})}\BibitemShut {NoStop}%
\bibitem [{\citenamefont {London}\ \emph {et~al.}(2014)\citenamefont {London},
  \citenamefont {Balasubramanian}, \citenamefont {Naydenov}, \citenamefont
  {McGuinness},\ and\ \citenamefont {Jelezko}}]{London2014StrongFields}%
  \BibitemOpen
  \bibfield  {author} {\bibinfo {author} {\bibfnamefont {P.}~\bibnamefont
  {London}}, \bibinfo {author} {\bibfnamefont {P.}~\bibnamefont
  {Balasubramanian}}, \bibinfo {author} {\bibfnamefont {B.}~\bibnamefont
  {Naydenov}}, \bibinfo {author} {\bibfnamefont {L.~P.}\ \bibnamefont
  {McGuinness}},\ and\ \bibinfo {author} {\bibfnamefont {F.}~\bibnamefont
  {Jelezko}},\ }\bibfield  {title} {\bibinfo {title} {Strong driving of a
  single spin using arbitrarily polarized fields},\ }\href
  {https://doi.org/10.1103/PhysRevA.90.012302} {\bibfield  {journal} {\bibinfo
  {journal} {Phys. Rev. A}\ }\textbf {\bibinfo {volume} {90}},\ \bibinfo
  {pages} {012302} (\bibinfo {year} {2014})}\BibitemShut {NoStop}%
\bibitem [{\citenamefont {Rao}\ and\ \citenamefont
  {Suter}(2017)}]{Rao2017NonlinearApproximation}%
  \BibitemOpen
  \bibfield  {author} {\bibinfo {author} {\bibfnamefont {K.~R.~K.}\
  \bibnamefont {Rao}}\ and\ \bibinfo {author} {\bibfnamefont {D.}~\bibnamefont
  {Suter}},\ }\bibfield  {title} {\bibinfo {title} {Nonlinear dynamics of a
  two-level system of a single spin driven beyond the rotating-wave
  approximation},\ }\href {https://doi.org/10.1103/PhysRevA.95.053804}
  {\bibfield  {journal} {\bibinfo  {journal} {Phys. Rev. A}\ }\textbf {\bibinfo
  {volume} {95}},\ \bibinfo {pages} {053804} (\bibinfo {year}
  {2017})}\BibitemShut {NoStop}%
\bibitem [{\citenamefont {Bidinosti}\ \emph {et~al.}(2022)\citenamefont
  {Bidinosti}, \citenamefont {Tastevin},\ and\ \citenamefont
  {Nacher}}]{Bidinosti2022GeneratingApproximation}%
  \BibitemOpen
  \bibfield  {author} {\bibinfo {author} {\bibfnamefont {C.~P.}\ \bibnamefont
  {Bidinosti}}, \bibinfo {author} {\bibfnamefont {G.}~\bibnamefont
  {Tastevin}},\ and\ \bibinfo {author} {\bibfnamefont {P.~J.}\ \bibnamefont
  {Nacher}},\ }\bibfield  {title} {\bibinfo {title} {Generating accurate tip
  angles for {NMR} outside the rotating-wave approximation},\ }\href
  {https://doi.org/10.1016/j.jmr.2022.107306} {\bibfield  {journal} {\bibinfo
  {journal} {J. Magn. Reson.}\ }\textbf {\bibinfo {volume} {345}},\ \bibinfo
  {pages} {107306} (\bibinfo {year} {2022})}\BibitemShut {NoStop}%
\bibitem [{\citenamefont {Bloch}\ and\ \citenamefont
  {Siegert}(1940)}]{Bloch1940MagneticFields}%
  \BibitemOpen
  \bibfield  {author} {\bibinfo {author} {\bibfnamefont {F.}~\bibnamefont
  {Bloch}}\ and\ \bibinfo {author} {\bibfnamefont {A.}~\bibnamefont
  {Siegert}},\ }\bibfield  {title} {\bibinfo {title} {Magnetic resonance for
  nonrotating fields},\ }\href {https://doi.org/10.1103/PhysRev.57.522}
  {\bibfield  {journal} {\bibinfo  {journal} {Phys. Rev.}\ }\textbf {\bibinfo
  {volume} {57}},\ \bibinfo {pages} {522} (\bibinfo {year} {1940})}\BibitemShut
  {NoStop}%
\bibitem [{\citenamefont {Zeuch}\ \emph {et~al.}(2020)\citenamefont {Zeuch},
  \citenamefont {Hassler}, \citenamefont {Slim},\ and\ \citenamefont
  {DiVincenzo}}]{Zeuch2020ExactApproximation}%
  \BibitemOpen
  \bibfield  {author} {\bibinfo {author} {\bibfnamefont {D.}~\bibnamefont
  {Zeuch}}, \bibinfo {author} {\bibfnamefont {F.}~\bibnamefont {Hassler}},
  \bibinfo {author} {\bibfnamefont {J.~J.}\ \bibnamefont {Slim}},\ and\
  \bibinfo {author} {\bibfnamefont {D.~P.}\ \bibnamefont {DiVincenzo}},\
  }\bibfield  {title} {\bibinfo {title} {Exact rotating wave approximation},\
  }\href {https://doi.org/10.1016/j.aop.2020.168327} {\bibfield  {journal}
  {\bibinfo  {journal} {Ann. Phys.}\ }\textbf {\bibinfo {volume} {423}},\
  \bibinfo {pages} {168327} (\bibinfo {year} {2020})}\BibitemShut {NoStop}%
\bibitem [{\citenamefont {Vandersypen}\ and\ \citenamefont
  {Chuang}(2005)}]{Vandersypen2005NMRComputation}%
  \BibitemOpen
  \bibfield  {author} {\bibinfo {author} {\bibfnamefont {L.~M.~K.}\
  \bibnamefont {Vandersypen}}\ and\ \bibinfo {author} {\bibfnamefont {I.~L.}\
  \bibnamefont {Chuang}},\ }\bibfield  {title} {\bibinfo {title} {{NMR
  techniques for quantum control and computation}},\ }\href
  {https://doi.org/10.1103/RevModPhys.76.1037} {\bibfield  {journal} {\bibinfo
  {journal} {Rev. Mod. Phys.}\ }\textbf {\bibinfo {volume} {76}},\ \bibinfo
  {pages} {1037} (\bibinfo {year} {2005})}\BibitemShut {NoStop}%
\bibitem [{\citenamefont {Bradley}\ \emph {et~al.}(2019)\citenamefont
  {Bradley}, \citenamefont {Randall}, \citenamefont {Abobeih}, \citenamefont
  {Berrevoets}, \citenamefont {Degen}, \citenamefont {Bakker}, \citenamefont
  {Markham}, \citenamefont {Twitchen},\ and\ \citenamefont
  {Taminiau}}]{Bradley2019AMinute}%
  \BibitemOpen
  \bibfield  {author} {\bibinfo {author} {\bibfnamefont {C.~E.}\ \bibnamefont
  {Bradley}}, \bibinfo {author} {\bibfnamefont {J.}~\bibnamefont {Randall}},
  \bibinfo {author} {\bibfnamefont {M.~H.}\ \bibnamefont {Abobeih}}, \bibinfo
  {author} {\bibfnamefont {R.~C.}\ \bibnamefont {Berrevoets}}, \bibinfo
  {author} {\bibfnamefont {M.~J.}\ \bibnamefont {Degen}}, \bibinfo {author}
  {\bibfnamefont {M.~A.}\ \bibnamefont {Bakker}}, \bibinfo {author}
  {\bibfnamefont {M.}~\bibnamefont {Markham}}, \bibinfo {author} {\bibfnamefont
  {D.~J.}\ \bibnamefont {Twitchen}},\ and\ \bibinfo {author} {\bibfnamefont
  {T.~H.}\ \bibnamefont {Taminiau}},\ }\bibfield  {title} {\bibinfo {title} {A
  ten-qubit solid-state spin register with quantum memory up to one minute},\
  }\href {https://doi.org/10.1103/PhysRevX.9.031045} {\bibfield  {journal}
  {\bibinfo  {journal} {Phys. Rev. X}\ }\textbf {\bibinfo {volume} {9}},\
  \bibinfo {pages} {031045} (\bibinfo {year} {2019})}\BibitemShut {NoStop}%
\bibitem [{\citenamefont {Fowler}\ \emph {et~al.}(2012)\citenamefont {Fowler},
  \citenamefont {Mariantoni}, \citenamefont {Martinis},\ and\ \citenamefont
  {Cleland}}]{Fowler2012SurfaceComputation}%
  \BibitemOpen
  \bibfield  {author} {\bibinfo {author} {\bibfnamefont {A.~G.}\ \bibnamefont
  {Fowler}}, \bibinfo {author} {\bibfnamefont {M.}~\bibnamefont {Mariantoni}},
  \bibinfo {author} {\bibfnamefont {J.~M.}\ \bibnamefont {Martinis}},\ and\
  \bibinfo {author} {\bibfnamefont {A.~N.}\ \bibnamefont {Cleland}},\
  }\bibfield  {title} {\bibinfo {title} {{Surface codes: Towards practical
  large-scale quantum computation}},\ }\href
  {https://doi.org/10.1103/PhysRevA.86.032324} {\bibfield  {journal} {\bibinfo
  {journal} {Phys. Rev. A}\ }\textbf {\bibinfo {volume} {86}},\ \bibinfo
  {pages} {032324} (\bibinfo {year} {2012})}\BibitemShut {NoStop}%
\bibitem [{\citenamefont {Barends}\ \emph {et~al.}(2014)\citenamefont
  {Barends}, \citenamefont {Kelly}, \citenamefont {Megrant}, \citenamefont
  {Veitia}, \citenamefont {Sank}, \citenamefont {Jeffrey}, \citenamefont
  {White}, \citenamefont {Mutus}, \citenamefont {Fowler}, \citenamefont
  {Campbell}, \citenamefont {Chen}, \citenamefont {Chen}, \citenamefont
  {Chiaro}, \citenamefont {Dunsworth}, \citenamefont {Neill}, \citenamefont
  {O’Malley}, \citenamefont {Roushan}, \citenamefont {Vainsencher},
  \citenamefont {Wenner}, \citenamefont {Korotkov}, \citenamefont {Cleland},\
  and\ \citenamefont {Martinis}}]{Barends2014SuperconductingTolerance}%
  \BibitemOpen
  \bibfield  {author} {\bibinfo {author} {\bibfnamefont {R.}~\bibnamefont
  {Barends}}, \bibinfo {author} {\bibfnamefont {J.}~\bibnamefont {Kelly}},
  \bibinfo {author} {\bibfnamefont {A.}~\bibnamefont {Megrant}}, \bibinfo
  {author} {\bibfnamefont {A.}~\bibnamefont {Veitia}}, \bibinfo {author}
  {\bibfnamefont {D.}~\bibnamefont {Sank}}, \bibinfo {author} {\bibfnamefont
  {E.}~\bibnamefont {Jeffrey}}, \bibinfo {author} {\bibfnamefont {T.~C.}\
  \bibnamefont {White}}, \bibinfo {author} {\bibfnamefont {J.}~\bibnamefont
  {Mutus}}, \bibinfo {author} {\bibfnamefont {A.~G.}\ \bibnamefont {Fowler}},
  \bibinfo {author} {\bibfnamefont {B.}~\bibnamefont {Campbell}}, \bibinfo
  {author} {\bibfnamefont {Y.}~\bibnamefont {Chen}}, \bibinfo {author}
  {\bibfnamefont {Z.}~\bibnamefont {Chen}}, \bibinfo {author} {\bibfnamefont
  {B.}~\bibnamefont {Chiaro}}, \bibinfo {author} {\bibfnamefont
  {A.}~\bibnamefont {Dunsworth}}, \bibinfo {author} {\bibfnamefont
  {C.}~\bibnamefont {Neill}}, \bibinfo {author} {\bibfnamefont
  {P.}~\bibnamefont {O’Malley}}, \bibinfo {author} {\bibfnamefont
  {P.}~\bibnamefont {Roushan}}, \bibinfo {author} {\bibfnamefont
  {A.}~\bibnamefont {Vainsencher}}, \bibinfo {author} {\bibfnamefont
  {J.}~\bibnamefont {Wenner}}, \bibinfo {author} {\bibfnamefont {A.~N.}\
  \bibnamefont {Korotkov}}, \bibinfo {author} {\bibfnamefont {A.~N.}\
  \bibnamefont {Cleland}},\ and\ \bibinfo {author} {\bibfnamefont {J.~M.}\
  \bibnamefont {Martinis}},\ }\bibfield  {title} {\bibinfo {title}
  {Superconducting quantum circuits at the surface code threshold for fault
  tolerance},\ }\href {https://doi.org/10.1038/nature13171} {\bibfield
  {journal} {\bibinfo  {journal} {Nature}\ }\textbf {\bibinfo {volume} {508}},\
  \bibinfo {pages} {500} (\bibinfo {year} {2014})}\BibitemShut {NoStop}%
\bibitem [{\citenamefont {A.~P.~Peirce}(1988)}]{Rabitz}%
  \BibitemOpen
  \bibfield  {author} {\bibinfo {author} {\bibfnamefont {H.~R.}\ \bibnamefont
  {A.~P.~Peirce}, \bibfnamefont {M.~A.~Dahleh}},\ }\bibfield  {title} {\bibinfo
  {title} {Optimal control of quantum-mechanical systems: Existence, numerical
  approximation, and application},\ }\href@noop {} {\bibfield  {journal}
  {\bibinfo  {journal} {Phys. Rev. A}\ }\textbf {\bibinfo {volume} {37}},\
  \bibinfo {pages} {4950} (\bibinfo {year} {1988})}\BibitemShut {NoStop}%
\bibitem [{\citenamefont {Kosloff}\ \emph {et~al.}(1989)\citenamefont
  {Kosloff}, \citenamefont {Rice}, \citenamefont {Gaspard}, \citenamefont
  {Tersigni},\ and\ \citenamefont {Tannor}}]{k67}%
  \BibitemOpen
  \bibfield  {author} {\bibinfo {author} {\bibfnamefont {R.}~\bibnamefont
  {Kosloff}}, \bibinfo {author} {\bibfnamefont {S.~A.}\ \bibnamefont {Rice}},
  \bibinfo {author} {\bibfnamefont {P.}~\bibnamefont {Gaspard}}, \bibinfo
  {author} {\bibfnamefont {S.}~\bibnamefont {Tersigni}},\ and\ \bibinfo
  {author} {\bibfnamefont {D.}~\bibnamefont {Tannor}},\ }\bibfield  {title}
  {\bibinfo {title} {Wavepacket dancing: Achieving chemical selectivity by
  shaping light pulses},\ }\href@noop {} {\bibfield  {journal} {\bibinfo
  {journal} {Chem. Phys.}\ }\textbf {\bibinfo {volume} {139}},\ \bibinfo
  {pages} {201} (\bibinfo {year} {1989})}\BibitemShut {NoStop}%
\bibitem [{\citenamefont {Palao}\ and\ \citenamefont {Kosloff}(2003)}]{k193}%
  \BibitemOpen
  \bibfield  {author} {\bibinfo {author} {\bibfnamefont {J.~P.}\ \bibnamefont
  {Palao}}\ and\ \bibinfo {author} {\bibfnamefont {R.}~\bibnamefont
  {Kosloff}},\ }\bibfield  {title} {\bibinfo {title} {Optimal control theory
  for unitary transformations},\ }\href@noop {} {\bibfield  {journal} {\bibinfo
   {journal} {Phys. Rev. A}\ }\textbf {\bibinfo {volume} {68}},\ \bibinfo
  {pages} {062308} (\bibinfo {year} {2003})}\BibitemShut {NoStop}%
\bibitem [{\citenamefont {Werschnik}\ and\ \citenamefont
  {Gross}(2007)}]{gross07}%
  \BibitemOpen
  \bibfield  {author} {\bibinfo {author} {\bibfnamefont {J.}~\bibnamefont
  {Werschnik}}\ and\ \bibinfo {author} {\bibfnamefont {E.~K.~U.}\ \bibnamefont
  {Gross}},\ }\bibfield  {title} {\bibinfo {title} {Quantum optimal control
  theory},\ }\href {https://doi.org/10.1088/0953-4075/40/18/r01} {\bibfield
  {journal} {\bibinfo  {journal} {J. Phys. B: At. Mol. Opt. Phys.}\ }\textbf
  {\bibinfo {volume} {40}},\ \bibinfo {pages} {R175} (\bibinfo {year}
  {2007})}\BibitemShut {NoStop}%
\bibitem [{\citenamefont {Schaefer}(2012)}]{Ido_thesis}%
  \BibitemOpen
  \bibfield  {author} {\bibinfo {author} {\bibfnamefont {I.}~\bibnamefont
  {Schaefer}},\ }\emph {\bibinfo {title} {Quantum Optimal Control Theory of
  Harmonic Generation}},\ \href {https://doi.org/10.48550/arXiv.1202.6520}
  {Master's thesis},\ \bibinfo  {school} {The Hebrew University of Jerusalem}
  (\bibinfo {year} {2012})\BibitemShut {NoStop}%
\end{thebibliography}%

\end{document}